\documentclass[
  journal=pas,
  manuscript=article,
  year=2025,
  volume=42,
  hyperref,
]{cup-journal}

\providecommand{\apjl}{ApJL}
\providecommand{\apjs}{ApJS}
\providecommand{\aap}{A\&A}

\makeatletter
\renewcommand*\cup@reference@code{%
  \RequirePackage[authoryear,round,sort]{natbib}%
}
\makeatother



\usepackage{amsmath}
\usepackage[nopatch]{microtype}
\usepackage{booktabs}
\usepackage{appendix}
\usepackage{xurl}

\title{Comparative analysis of BL Lacertae in flaring and non-flaring states: timing and spectral studies}

\author{A. Priyana Noel}
\affiliation{Astronomical Observatory, Jagiellonian University, Orla 171, PL-30-244, Krakow, Poland}
\email[F. Author]{anoel@oa.uj.edu.pl}

\author{ Alicja Wierzcholska}
\affiliation{Institute of Nuclear Physics, Polish Academy of Sciences, ul. Radzikowskiego 152, 31-342 Kraków, Poland}

\author{Raj Prince}
\affiliation{Department of Physics, Institute of Science, Banaras Hindu University, Varanasi-221005, India}
\alsoaffiliation{
Center for Theoretical Physics, Polish Academy of Sciences, Al.Lotnikov 32/46, PL-02-668 Warsaw, Poland}


\begin{document}

\begin{abstract}
Context: BL Lacertae is a blazar known for its high flux variability and occasional broadband flares, the origins of which remain unknown. BL Lacertae was found to be in an extended flaring state in July 2020 which continued until the end of 2021. \\
Aims: The long-term flaring activity makes it an ideal candidate to study its spectral and temporal properties during different flux states. This study explores the X-ray temporal and spectral variability of BL Lacertae. \\
Methods: We analysed five observations of BL Lacertae with the XMM-Newton EPIC instrument taken up to the end of 2021. Temporal properties were investigated using the fractional variability method, minimum variability timescale, and the discrete correlation function. Detailed spectral modeling was performed on the two most variable observations, including an investigation of correlations between the soft (0.3 - 2.0 keV) and hard (2.0 - 10.0 keV) energy bands. \\
Results: Out of five observations, two observations were found to be highly variable with F$_{var}$=19.16$\pm$0.32 and 6.27$\pm$0.43. The observation taken in 2021 corresponds to the highest flux state. The shortest variability timescale in the 0.3-10 keV band is estimated as 1.24 ks. Assuming the X-ray emission is dominated by the synchrotron process, this variability timescale constrains the size of the emission region. Under the assumption of equipartition between the magnetic field and radiating particles, this implies a magnetic field strength of B $\approx$ 0.4G. The spectral analysis reveals a softer-when-brighter trend, which is commonly seen in blazars. We modeled the X-ray spectra with single power-law, log-parabola, and broken power-law models. In most cases, a broken power-law provided the best fit based on corrected Akaike Information Criterion (AICc) statistics, and a strong correlation was observed between the break energy and the source flux. When a thermal blackbody component was added to the model, its temperature also showed a positive correlation with flux in some observations. \\
Conclusions: Our work indicates the complex spectral evolution of BL Lacertae during this flare. The spectral break, interpreted as the cooling break within the synchrotron component, shifts to higher energies with increasing flux. The source consistently displayed softer-when-brighter behavior. In only one observation were the soft and hard bands found to be significantly correlated. The data suggest a scenario where the peak of the synchrotron emission moves into or across the X-ray band as the source brightens.
\end{abstract}

\section{Introduction}
\noindent
Blazars are a category of radio-loud active galactic nuclei with relativistic jets facing the Earth. Blazars emit in the entire observable electromagnetic spectrum, displaying variability over flux and spectra with time. These are classified as BL Lacertae (BL Lac) and flat spectrum radio quasars (FSRQs), depending on the absence or presence of the emission lines. FSRQs show strong and broad emission lines with an equivalent width (EW) of greater than 5{\AA} and BL Lac objects have a weak or featureless spectrum with EW less than 5{\AA}. Blazars display variability in flux across different timescales ranging from minutes to hours and even years.\\
The multi-wavelength spectral energy distribution (SED) of blazars in $\nu$-f$\nu$ representation depicts broadband double-humped structures \citep[e.g.,][]{1997A&A...327...61G}.     The low-energy hump, spanning the radio to X-ray bands, is produced by synchrotron emission from relativistic, non-thermal charged particles. The high-energy peak, extending from X-ray to TeV energies, originates from either leptonic or hadronic processes.
In leptonic models, this high-energy emission results from relativistic electrons upscattering photons via the inverse Compton (IC) mechanism. The target photons are either the synchrotron photons themselves (Synchrotron-Self Compton, or SSC) or photons from an external source like the broad-line region, dusty torus, or accretion disk (External Compton, or EC). \citep[e.g.,][]{1998A&A...333..452K,2010ApJ...718..279G}.
The hadronic model attributes the emission to the proton synchrotron emission and synchrotron and Compton emission from the decay of charged pions or the decay of neutral pions \citep[e.g.,][]{2003APh....18..593M}. 
Blazars with peak of the synchrotron hump lying below $10^{14}$ Hz are called low-energy peaked blazars (LBLs) and those with peaks in between $10^{14}$ Hz  and $10^{15}$ Hz are called intermediate-energy peaked blazars (IBLs). High-energy peaked blazars (HBLs) are blazars with the synchrotron peak lying above $10^{15}$ Hz.  
The SED of blazars in X-ray helps to understand both the flux and spectral variability in the synchrotron and inverse Compton emission. For LBLs, the X-ray emission consists of inverse Compton emission in the leptonic scenario. The IBLs portray both the synchrotron and inverse Compton emission in X-ray and the spectra can be used to estimate the energy at which the upturn from synchrotron to inverse Compton occurs. The SED of HBLs shows the synchrotron part of the emission. Among many \citet{10.1093/mnras/stw095} and \citet{2006A&A...457..133F}, worked on the X-ray spectra of blazars to study the upturn from the synchrotron emission to inverse Compton emission. \\
The models commonly used to describe the X-ray spectra of BL Lac objects include the single power-law \citep{1990ApJ...360..396W}, broken power-law \citep{1996ApJ...463..444S}, and log-parabola \citep{1996ApJ...463..555I} models. The log-parabola model effectively describes the X-ray emission from a large fraction of BL Lac objects \citep{2021MNRAS.507.5690G, 2005A&A...433.1163D, 1998ApJ...509..608T}. The double power-law model provides a better representation of the X-ray emission from blazars than the single power-law when both synchrotron and inverse Compton components contribute to the X-ray emission. Both double power-law model and log-parabola model help to identify the energy at which the spectral upturn occurs \citep{2006A&A...457..133F}.
As noted by \citet{2005A&A...433.1163D}, HBLs typically exhibit an X-ray spectral index between 3 and 1.8, while for other blazars it can be lower than 1.8. The hardness of the photons also varies with flux variability, with \citet{2001A&A...365L.162B} reporting that flux variability is higher at higher energies. \\
One of the most common blazar among many that has been studied for both synchrotron and inverse Compton emission in X-ray is BL Lacertae (also known as BL Lac but for clarity, we will refer to it by full name). 
However, BL Lacertae was referred as an LBL by \citet{2018A&A...620A.185N} when the logarithmic value of synchrotron peak frequency in Hertz was found 13.99. This object displays complex behavior with changing spectral behavior \citep{2021MNRAS.507.5602P}. BL Lacertae like other blazars exhibits timing and spectral variability and has been monitored using various multiwavelength campaigns. 

\subsection{Previous X-ray results of BL Lacertae}
Before BL Lacertae flared in X-rays (in the 0.3–10.0 keV band examined in this work) during 2020–2021, the two most widely studied X-ray flares of BL Lacertae were those of 2001 and 2011–2012 \citep{2021MNRAS.507.5602P}.
\citet{2003A&A...408..479R} examined X-ray observations of BL Lacertae obtained with BeppoSAX and RXTE-PCA in July and October 2000. The BeppoSAX data showed a low-flux state on July 26–27, followed by a high-flux state during October 31 to November 2, 2000. In July, when the source was in a steady state, it exhibited a hard X-ray spectrum. The flux in the 0.7–2.0 keV energy range varied on timescales of a few hours, while the flux in the 2.0–10.0 keV range remained nearly constant. In October 2000, a transition from a soft to hard spectrum was observed in the 10–20 keV energy range.
  \\
An extensive multiwavelength analysis of BL Lacertae was conducted by \citet{2003ApJ...596..847B} for observations spanning from mid-May 2000 to the end of that year. The group collected data in the radio, optical, X-ray, and very high energy (VHE) bands, presenting results on flux and spectral variability as well as broadband SEDs. This study showed that the source was quiescent before September 2000 and then entered a prolonged flaring state. During the high-flux state, the X-ray band exhibited increased variability on timescales of less than a few hours. Evidence was found for the contribution of synchrotron emission up to 10 keV during the flaring state.
An examination of the hardness–intensity relation using BeppoSAX data revealed that in the soft X-ray band (0.5–2 keV), the spectrum softened as the flux increased, while in the medium X-ray band (2–4 keV), the usual "harder-when-brighter" behavior was confirmed. The study also reported spectral softening with increasing flux in RXTE-PCA observations. Additionally, they found that optical variability traced the soft X-ray variability, but with a time lag of 4–5 hours.
\\
Earlier \textit{XMM-Newton} observations of BL Lacertae were analysed by \citet{2009A&A...507..769R}, who reported that the X-ray spectra exhibited contributions from both synchrotron and inverse Compton emission. Previous studies on BL Lacertae have shown moderate intraday variability in optical multiwavelength observations (e.g., \citet{2017Galax...5...94G}).
\citet{2018A&A...619A..93B} analysed X-ray observations of BL Lacertae taken with \textit{NuSTAR} in December 2012 and found no correlation between hardness and flux during the observation. \citet{10.1093/mnras/stw095} analysed \textit{Swift}-XRT observations of BL Lacertae and found that both the power-law and log-parabola models provided good fits to the observations. Their results showed different values of the spectral index and curvature depending on whether Galactic absorption was allowed to vary or kept fixed.   \\
The source entered a flaring state in 2020, and several studies have examined observations taken during this period across multiple wavelengths. \citet{2022MNRAS.513.4645S} conducted an extensive analysis of BL Lacertae using observations from 2008 to 2020, studying the flares that occurred in 2012 and 2020 and explaining the complex emission behavior of the source. \citet{2021MNRAS.507.5602P} investigated the broadband spectral evolution of BL Lacertae during its 2020 flaring period, analyzing the October 2020 flare across multiple wavebands. Their studies suggested that BL Lacertae exhibited HBL-like behavior during this flaring state.

\subsection{BL Lacertae enters in a flaring state}
BL Lacertae flared in the optical, X-ray, and gamma-ray bands in 2020. Gamma-ray flaring was observed with \textit{Fermi}-LAT, coinciding with the optical flare \citep{2020ATel13933....1C}. 
MAGIC detected VHE flaring on 19 September 2020 \citep{2020ATel13963....1B}, following the flaring in GeV energy ranges reported by \textit{Fermi}-LAT, which had been detected on 19 August 2020.
\citet{2020ATel14065....1D} reported that the source reached its historical flux maximum in a \textit{Swift}-XRT observation on 6 October 2020, which was contemporaneous with gamma-ray flaring observed by \textit{Fermi}-LAT. On 11–12 October 2020, the source was found to be flaring again with \textit{NuSTAR}, reaching a flux as high as $3.44$ $\times$ $10^{-10}$ erg cm$^{-2}$ s$^{-1}$ in the 0.3–10 keV band \citet{2022MNRAS.509...52D}.
The source reached another maximum on 18 January 2021 in the optical band, accompanied by gamma-ray flaring detected by \textit{Fermi}-LAT \citep{2021ATel14330....1C}. The flaring state continued throughout 2021, and until August 2021, BL Lacertae was reported to be flaring in the optical, X-ray, and gamma-ray bands. \citet{2021MNRAS.507.5602P} reported continued X-ray flaring activity from BL Lacertae, observed with \textit{Swift}-XRT from 8 to 11 July 2021 \citep{2021ATel14774....1P}.

\section{Data selection and reduction}
The observations for the study of intraday variability and spectral analysis of BL Lacertae have been taken from the \textit{XMM-Newton} satellite.

\subsection{XMM-Newton}
\textit{XMM-Newton} satellite was launched in 1999. The instruments onboard \textit{XMM-Newton} are -- three European Photon Imaging Cameras (EPIC), two Reflection Grating Spectrometers (RGS), and one Optical Monitor telescope. \\
EPIC is an X-ray instrument with three CCDs - one of pn type and other two of MOS type. Each of the CCD has an effective area of approximately 1500 $cm^2$ with a large field of view of 30' diameter. EPIC can observe in the photon energy range of 0.15 -- 15 keV.
\\
The time resolution of the CCDs of EPIC-pn telescope in timing and burst mode is 30 $\mu$s which is higher than the time resolution of EPIC MOS telescope. Only 40\% of the flux is incident on the MOS cameras because MOS cameras are behind the grating of RGS. Thus we have used X-ray observations taken from EPIC-pn instrument which captures the total incident flux.

\subsection{Data selection}
The archival observations of BL Lacertae taken using the PN detector of EPIC-pn instrument onboard \textit{XMM-Newton} have been used for the computation of variability.  There are five observations \footnote{(at the time of beginning the project)} of BL lacerate taken from July 2007 till July 2021 which have been downloaded from \textit{XMM-Newton} Science Archive \footnote{\url{http://nxsa.esac.esa.int/nxsa-web/\#search}}. Here we used the PN observations  because of the higher sensitivity and lower photon pile-up effects of the PN detector as compared to the MOS detector of EPIC. All five observations have an observing time of more than 10 ks which makes all of these suitable for time-dependent spectral variability studies. The five observations that have been selected are given in table \ref{observation_table} along with time and duration of observation. \\
The observation taken in July 2021 coincides with the source being reported as flaring in X-ray by \citet{2021ATel14774....1P}. The other four EPIC-pn observations are from the non-flaring state of BL Lacertae. \\

 \subsection{Data reduction and analysis}
 The observations were reduced to get spectrum and light curves using the Scientific Analysis System (SAS) software (SAS 21.0.)\footnote{https://www.cosmos.esa.int/web/xmm-newton/sas-threads}.  Observation Data Files (ODFs) downloaded from the \textit{XMM-Newton} data archive consist of the instrument science data, housekeeping and auxiliary files. It also includes the Current Calibration Files (CCFs) which are needed for data calibration. Tasks \textit{cifbuild} and \textit{odfingest} are used to obtain the list of calibration files and a summary file of all files in ODF set. All the tasks needed to produce an event list from PN data are compiled in one \textit{epchain} task. Task \textit{evselect} was used to create lightcurve and spectrum from a selected region. \\
 Pile-up correction is another important part of analyzing the observation. The pile-up effect occurs when more than one photon falls on a pixel and is counted as one photon of added energy of the striking photons. To correct the pile-up effect, the model distribution of a single photon event is compared to a double photon event. The overlap of the two distributions signifies the absence of any significant pile-up. To correct this effect in observation, the pile-up affected region is excised and only an annular region is selected. The task \textit{epatplot} has been used for this correction\footnote{\url{https://heasarc.gsfc.nasa.gov/docs/xmm/sas/USG/epicpileup.html}}.
 This effect was observed only in one observation, 0501660201, which was corrected by using the standard procedure of choosing an annular region of source while for all other observations, a circular region was chosen. Source region selected from observation 0501660201 was an annulur region lying between the concentric circles with radius of 230 arcseconds and 524 arcseconds centered at the source. The source region selected for observations 0501660301 and 0501660401 was a cicular region of radius 656 and 360 arcsecond centered at the source.  The background region was selected from a circular region on the same CCD outside the source region. For observations taken in timing mode, source and background spectra were extracted from 1D RAWX strips, not 2D regions. Dedicated Timing mode RMFs and ARFs were generated, correctly accounting for the geometry and fast readout. Timing mode inherently mitigates pile-up, and we confirmed it was negligible. Light curves in soft energy bands were carefully inspected, and no significant soft flares were detected requiring additional filtering beyond standard particle background screening. For the two Timing Mode observations (0504370401 and 0891800501), the source region was extracted from detector columns 37–39, centered on the source. The background region was extracted from columns 30–33, outside of the source region. \\
In the analysis we use the data from the limited energy range, from 0.3 keV to 10.0 keV, as there is known detector noise below 0.3 keV and  the measurements above 10.0 keV are subject to a high proton background caused by the solar activity \citep{Bulbul_2020, 2010ApJ...718..279G}. We monitored such effects by checking the derived light curves in this energy range. \\
The light curves and spectrum were obtained for further analysis. The time binning for the light curves was chosen on an observation-by-observation basis to optimally resolve the temporal structure of the source variability. The bin time of light curves was obtained by generating multiple light curves of different bin size and inspecting them. The time bin for each observation was chosen depending on the resolution that displayed characteristic features such as rise and decay of flares and the bin duration had to be long enough to ensure sufficient signal to noise ratio. The spectra were binned to minimum 25 counts per bin to ensure significant signal to noise ratio.

\section{Methodology -- Analysis techniques}
Various methods have been used to analyse the flux and spectral variability of X-ray observations of BL Lacertae taken with the PN instrument onboard the \textit{XMM-Newton} satellite.

\subsection{Flux variability analysis}
The method of excess variance and fractional variability amplitude \citep{2003MNRAS.345.1271V} was used to test the variability of observations as described in \citet{2022ApJS..262....4N}. Fractional root mean square variability amplitude ($F_{var}$) is mathematically described as:
\begin{equation}
F_{var}=\sqrt{\frac{S^{2}-{\overline{\sigma^{2}_{err}}}}{\overline{x}^2}}
\end{equation} 
where S$^2$ is the variance and $\sigma_{err, i}$ is the corresponding uncertainty associated with every flux point.
The uncertainty of F$_{var}$ is given by
\begin{equation}
err(F_{var})=\sqrt{{\left(\sqrt{\frac{1}{2N}}\frac{\overline{\sigma^2_{err}}}{{\overline{x}^2}F_{var}}\right)}^2+{\left(\sqrt{\frac{\overline{\sigma^2_{err}}}{N}}\frac{1}{\overline{x}}\right)}^2}
\end{equation}
\\
F$_{var}$ along with $err(F_{var})$ serves as a quantification of variability.
The variable observations were further analysed for time-dependent spectral analysis which is further explained in the following sections.

\subsection{Hardness ratio analysis}
The hardness ratio (HR) is used to study the spectral states of the source and to compare photon hardness and variability between flaring and non-flaring states. HR analysis is advantageous as it does not depend on predefined models and allows the use of small time bins with low photon counts, making it efficient for studying spectral variability.
The variable observations were split into energy ranges of 0.3–2.0 keV (soft) and 2.0–10.0 keV (hard), using the standard 2.0 keV division to separate the light curve into soft and hard bands \citep{2018MNRAS.481.3563P, 2022ApJS..262....4N}. To calculate the hardness ratio, same method as \citet{2006ApJ...652..610P} was used : 
\begin{equation}
    HR = \frac{H}{S}
\end{equation}
where H is the flux in the hard energy band and S is the flux in the soft energy band in the analysed time bin. To derive and plot the hardness ratio against intensity, \textit{lcurve} task of the XRONOS program of HEASARC has been used.

\subsection{Discrete correlation function}
The soft and hard light curves of variable observations were analysed to study the correlation between each other using the method of discrete correlation function (DCF) \citep{1988ApJ...333..646E, 2022ApJS..262....4N}. Monte Carlo simulation was performed on artifically generated 1000 light curves to assess the significance of the DCF peaks from which we estimated the 1$\sigma$, 2$\sigma$, and 3$\sigma$ confidence bands. The significance of the observed DCF peak was determined relative to these confidence levels. \\
To evaluate the time lag, the maxima of DCF was fitted with a Gaussian model \citep{2019ApJ...884..125Z}. The Gaussian distribution is given by:
\begin{equation}
\large
        DCF(\tau) = A e^{-(\tau- \mu)/2\sigma^2} 
\end{equation}
where A is the maximum DCF value and $\sigma$ is the fitted width of the Gaussian function. A limited range of time $\tau$ is selected for each DCF function such that the DCF maxima lie in this range. The time lag is given by $\mu$ which is the difference between the time at which the DCF maxima lies and the center of time range $\tau$.

\subsection{Spectral analysis}
The spectral analysis of all five observations is done to understand the complex behavior of BL Lacertae. For this the spectra, in the energy range of 0.3 - 10.0 keV, were obtained using \textit{evselect}, a function to create spectra in the SAS software. 
The spectral fitting was done using XSPEC v12.12.0 which is a spectral fitting package included in the HEASoft software. 
\\
The models used for fitting are power-law (PL), logparabola (LP), broken power-law (BP), double power-law (DP), which is a combination of two power-law, and blackbody models.Each of the models used include galactic absorption.\footnote{\url{https://heasarc.gsfc.nasa.gov/xanadu/xspec/manual/node130.html}} 
The double powerlaw model used here is a phenomenological model describing the spectral shape and not a physical model.  A log-parabola model was employed to characterize both the spectral slope and its curvature. We note that while a log-parabola is not the ideal tool for a simple hardness analysis, its parameters provide a robust way to track spectral changes when significant curvature is present and power-law model may fail to adequately fit the spectrum.
\\
 In all logparabola models, the pivot energy (pivotE) was fixed at the standard value of 1 keV \citep{2023ApJ...959...61E, 2025A&A...693A.299W}. This conventional choice provides a stable reference point for model normalization and facilitates comparison with other studies \citep{2021MNRAS.504.5575K}.
\\
Each of these models was tested using the $\chi^{2}$ method of fitting and compared to each other for best fit using Akaike Information Criterion (AIC) \citep{2021ApJ...914...46G}. To ensure that the analysis is not influenced by small sample-bin sizes, the AIC has been adjusted accordingly and is given by AIC + $\frac{2k(k+1)}{n-k-1}$ where $n$ is the sample size and $k$ is the number of free parameters in a model. This corrected AIC is referred as AICc in the paper. \\
Galactic absorption (n$_H$) has been taken into account for BL Lacertae because interstellar absorption is observed along with a molecular cloud in the line of sight of BL Lacertae \citep{2002A&A...383..763R}. Previous studies on BL Lacertae have used varied values of n$_H$. \citet{2002A&A...383..763R} attempted to fix n$_H$ to the Galactic hydrogen column density measured at 21 cm, which is 2 $\times$ $10^{21}$ $cm^{2}$. Earlier, \citet{1993A&A...276L..33L} reported the presence of a molecular cloud along the line of sight to BL Lacertae. To account for molecular Galactic absorption, \citet{2009A&A...507..769R} froze n$_H$ at 3.4 $\times$ $10^{21}$ $cm^{2}$. However, the residuals remained high, so this value was not adopted.
Similarly, the value of 1.7 $\times$ $10^{21}$ $cm^{2}$, obtained from the HI4PI full-sky H I survey \citep{2016A&A...594A.116H}, did not fit our data well. Using this HI4PI value resulted in a reduced $\chi^{2}$ greater than 1, indicating a poor fit to our observations. The value of 2.7 $\times$ $10^{21}$ $cm^{2}$ as used by \citet{2022MNRAS.509...52D} was used for analysis.     This two-step approach successfully identified the best-fit continuum model for four of the five observations. This model was then used to investigate the absorption by allowing the nH parameter to vary freely in a subsequent fit.

\subsection{Interval-wise spectral analysis}
Of the two observations, only those exhibiting extreme variability were selected. The light curve of each variable observation was divided into time intervals based on lower and higher photon counts, as shown in figure \ref{LC_var}. The spectrum for each interval was then generated separately. This approach, used by \citet{2006A&A...457..133F} for the IBL source S5 0716+71, is effective for studying spectral variations on intraday timescales.
\\
Of the two variable observations, the observation 0891800501 (observed on 09-07-2021) corresponds to a flaring state, while 0504370401 (observed on 16-05-2008) corresponds to a non-flaring state. Thus, this interval-wise spectral analysis is useful for comparing observations at different flux states while also examining how the spectra change with flux on short timescales within an observation. \\
The broken power-law fitting was used to determine the break energy where the spectrum shows an upturn and the photon indices of the two power-laws change. Assuming the lower-energy power-law arises from synchrotron emission, the synchrotron contribution \citep{2006A&A...457..133F} in each spectrum was estimated by calculating the flux below the break energy and expressing it as a percentage of the total flux in the 0.3–10.0 keV energy range.

\section{Results}
\subsection{Flux variability}
The results of $F_{var}$ are given in table \ref{observation_table} along with the observation details. Out of 5 total observations of BL Lacertae taken with \textit{XMM-Newton} until August 2021, only two observations, 0504370401 (date 16-05-2008) and 0891800501 (date 09-07-2021) were found to be extremely variable. The observations that we find non-variable here were also analysed by \citet{2009A&A...507..769R} and they found the same non-variable behavior. The observation ID 0504370401 (date 16-05-2008) is the observation with highest variability out of five with $F_{var}$ value of 19.16\%. The observations 0501660201 (date 10-07-2007) and 0501660301 (date 05-12-2007) have extremely low levels of variability with $F_{var}$ value lesser or almost equal to 3 times the $\sigma_{err}$. 
The observation 05016601401 (08-01-2008) has an undefined $F_{var}$  and therefore variability values are not listed in table \ref{observation_table}. This can occur if the source shows very low intrinsic variability during the observation or due to low photon count statistics. However, this observation has a count rate of $3.42 \pm 0.05$ counts/sec , indicating that the photon statistics are sufficient, and the undefined $F_{var}$ is likely due to a lack of significant variability during this observation.

\begin{table*}[htbp]
\caption{Summary of observations taken from EPIC-PN instrument of \textit{XMM-Newton}. Photon counts is the average photon count of an observation. Here observation is abbreviated as \textit{obs}.}
\label{observation_table}
\centering
\begin{tabular}{cccccccccc}\hline \hline
Obs. ID    & Date of Obs. & Window mode & Obs. duration  & Pile up &  Photon counts & $F_{\rm var}$ & $\tau_{\rm var}$ \\
           & dd-mm-yyyy   &            &            (ks) &   &   counts s$^{-1}$ &    &   (s)\\\hline
$0501660201$ & $10-07-2007$ & $Small$  & $18.9$ & $yes$ & $4.01 \pm 0.05$ & $4.06\pm1.08$ & $2356.05 \pm 0.17$ \\
$0501660301$ & $05-12-2007$ & $Small$  & $19.7$ & $no$ & $3.67 \pm 0.03$ & $2.78\pm0.65$ & $3408.73 \pm 0.15$\\
$0501660401$ & $08-01-2008$ & $Small$  & $22.8$ & $no$ & $3.42 \pm 0.05$ & $-$ & $-$ \\
$0504370401$ & $16-05-2008$ & $Timing$  & $133.9$ & $no$ & $2.23 \pm 0.03$ & $19.16\pm0.32$ & $1279.67 \pm 0.09$\\
$0891800501$ & $09-07-2021$ & $Timing$ & $19.6$ & $no$ & $5.28 \pm 0.03$ & $6.27\pm0.43$ & $4352.53 \pm 0.11$\\
\hline
\end{tabular}
\end{table*}

\begin{figure*}[htbp]
    \centering
    \begin{tabular}{lll}
(a) & (b) & (c) \\
\includegraphics[width=50mm, height=50mm]{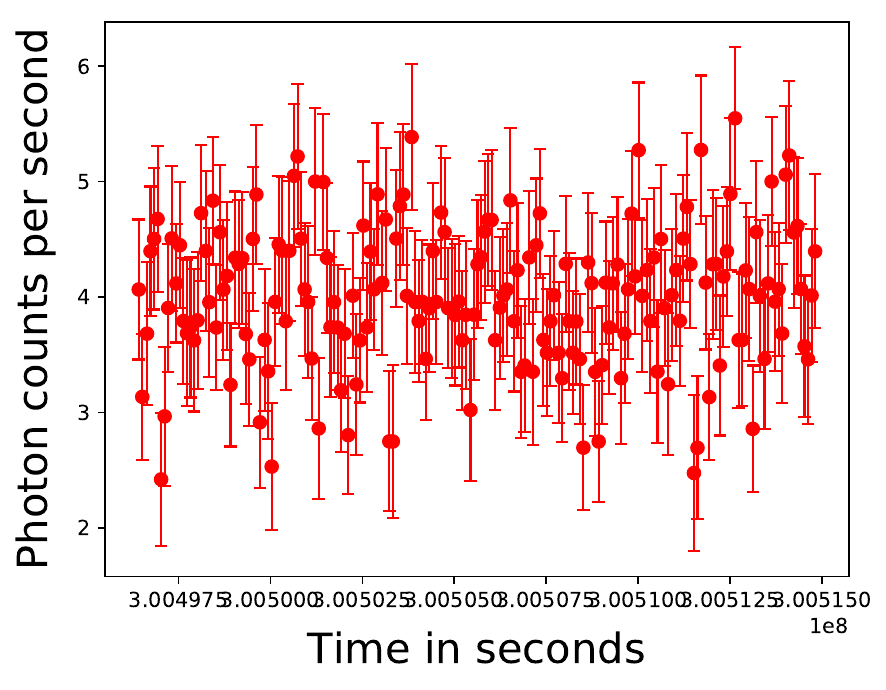} &
\includegraphics[width=50mm, height=50mm]{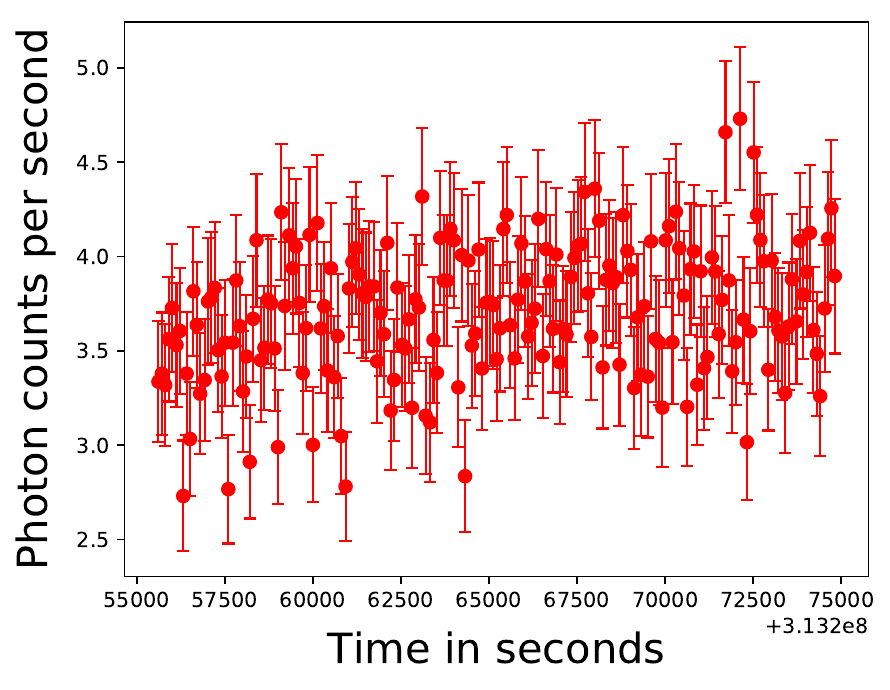} &
\includegraphics[width=50mm, height=50mm]{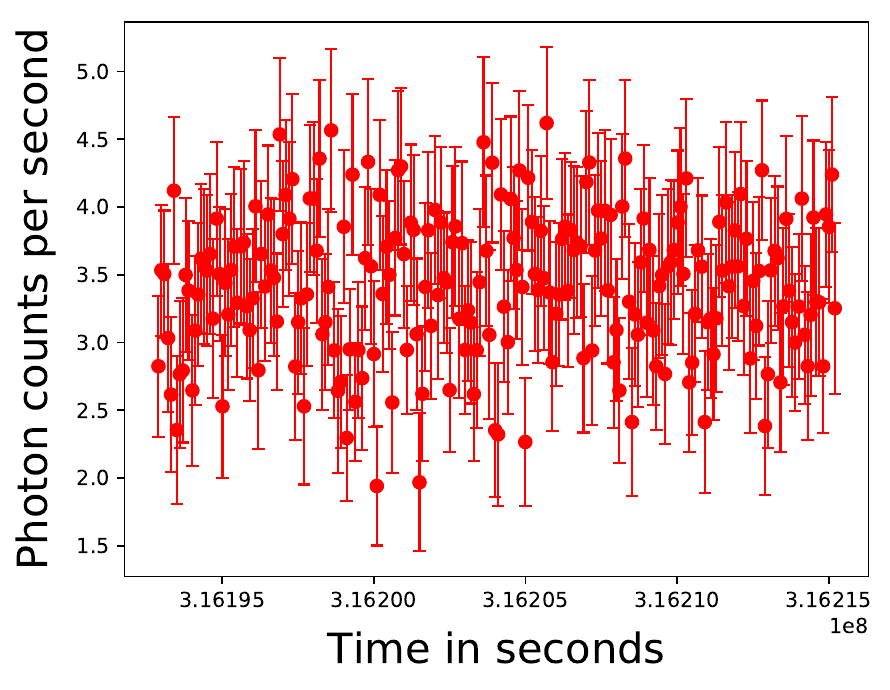} \\    
\end{tabular}
\caption{Light curves of the non-variable observations each with binning of 100 s (a) 0501660201, (b) 0501660301, (c) 0501660401.}
\label{lc_NV}
\end{figure*}

\begin{figure*}[htbp]
    \centering
    \begin{tabular}{cc}
(a)
\includegraphics[width=55mm, height=55mm]{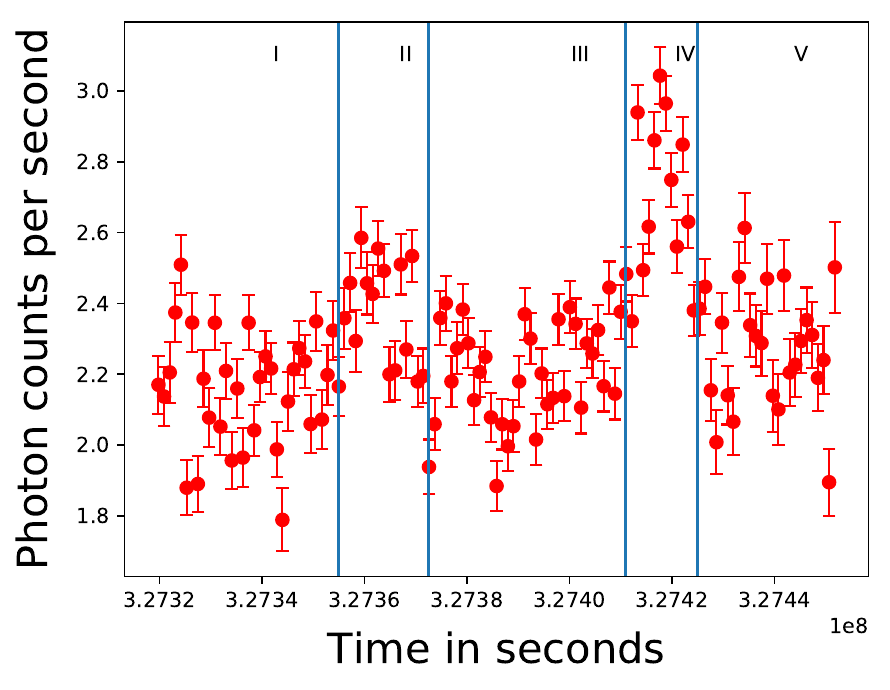} &
(b)
\includegraphics[width=55mm, height=55mm]{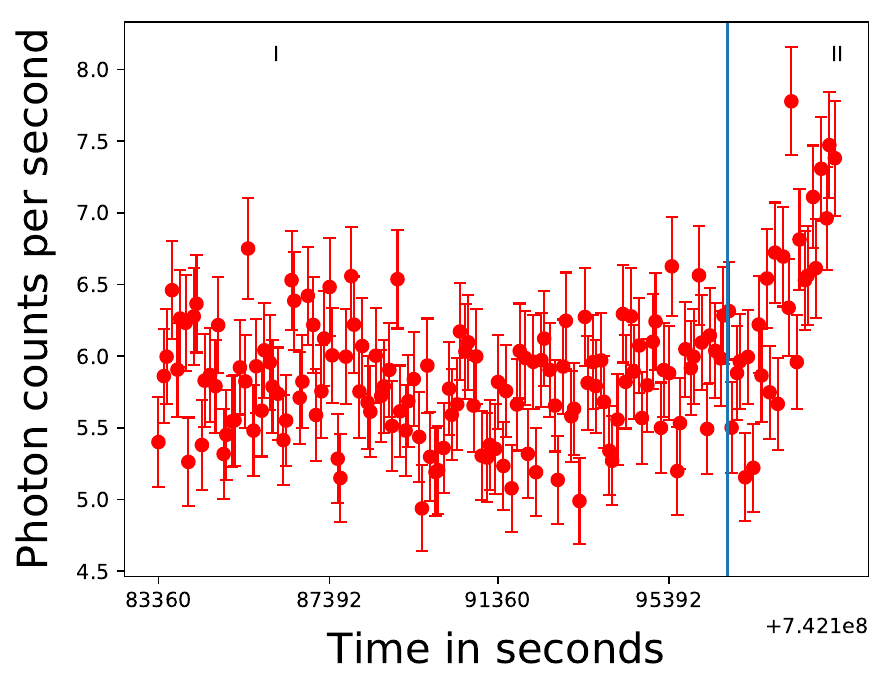} \\    
 
\end{tabular}
\caption{Light curves of the variable observations (a) 0504370401 with bin time of 1100 s, (b) 0891800501 with bin time of 100 s.}
\label{LC_var}
\end{figure*}

The highest variable observation also happens to be the observation with the longest duration i.e. 133.9 ks.
The other variable observation that was taken on 9 July 2021, which is 0891800501, coincides with the flaring period of BL Lacertae which started in 2020 and lasted over 2021. The light curves of the non-variable and the variable observations are shown in figure \ref{lc_NV} and figure \ref{LC_var} respectively. From the visual inspection as well, it is somewhat clear that only two observations shown in figure \ref{LC_var} display variability. Also from the light curves, it is clear that the observation taken in 2021 which coincides with the flaring stage of the source displays the highest flux count out of the five observations. Later we see a comparison between the spectra of the source in different variable regions of the variable observation. Also, the analysis of hardness ratio and lag in emission is studied of flaring (0891800501; 09-07-2021) and non-flaring observation (0504370401; 16-05-2008).

\subsection{Characteristic time scale}
The timescale of variability is an important parameter that can help us to derive some of the jet parameters such as the size of the emission region. We estimated the variability timescale for all the \textit{XMM-Newton} EPIC-PN observations and the formula used is:
\begin{equation}
    F(t_2) = F(t_1) \times 2^{(t_2 - t_1)/T_d}
\end{equation}
Here $F(t_1$) and $F(t_2$) are the observed fluxes at time $t_1$ and $t_2$, respectively. $T_d$ is the flux doubling/halving or variability timescale. From the variability estimate, we see that the observations 0504370401 (date 16-05-2008) and 0891800501 (date 09-07-2021) are more variable and hence can probe the smaller timescale. The variability timescale estimated for these two observations is 1.28 ks or 21.33 minutes and 4.35 ks or 1.2 hours, respectively. A minute and hour scale of variability in X-ray is common in blazar which puts tight constraints on the size of the emission region.

BL Lacertae is a complex source that sometimes behaves as an LBL type source and sometimes like an HBL type \citep{2021MNRAS.507.5602P} which suggests the synchrotron emission can produce the x-ray emission and hence the x-ray decay timescale can be linked with the radiation cooling or synchrotron timescale \citep{2001ApJS..132..377H,2002ApJ...572..392B,2014A&ARv..22...72U}.
 
The synchrotron cooling timescale can be defined as \citep{Rybicki1979},
\begin{equation}
    t_{cool} \simeq 7.74\times10^8  \frac{(1+z)}{\delta}B^{-2}\gamma^{-1} ~{\rm sec}.
\end{equation}
Where $B$ is the strength of the magnetic field in Gauss and $t_{\rm cool}$ is the synchrotron cooling timescale in seconds. Following \citep{Rybicki1979}, we can also derive the characteristic frequency of the electron population, 
responsible for the synchrotron emission at the SED peak,
\begin{equation} \label{eq:4}
    \nu_{ch,e} = 4.2\times10^6 \frac{\delta}{(1+z)}B \gamma^{2} ~{\rm Hz}.
\end{equation}
Using the above two equations, we can eliminate the $\gamma$ since it changes with different states and can derive a single equation given as,
\begin{equation}
    B^3\delta \simeq 2.5 (1+z)(\nu_{ch,e}/10^{18})^{-1} \tau_d^{-2}.
\end{equation}
where $\tau_d$ is in ks. Using the above equation, we derive the magnetic field strength for the Doppler factor (obtained from \citet{2021MNRAS.504.1427A, 2021MNRAS.507.5602P}), $\delta$ = 20, and a variability time scale of 1.28 ksec, as  0.4 G. For a range of Doppler factor between 10 and 30, magnetic field strength takes values between 0.5 G and 0.35 G.

\subsection{Spectral variability}
\begin{figure}
    \includegraphics[width=100mm,height=80mm]{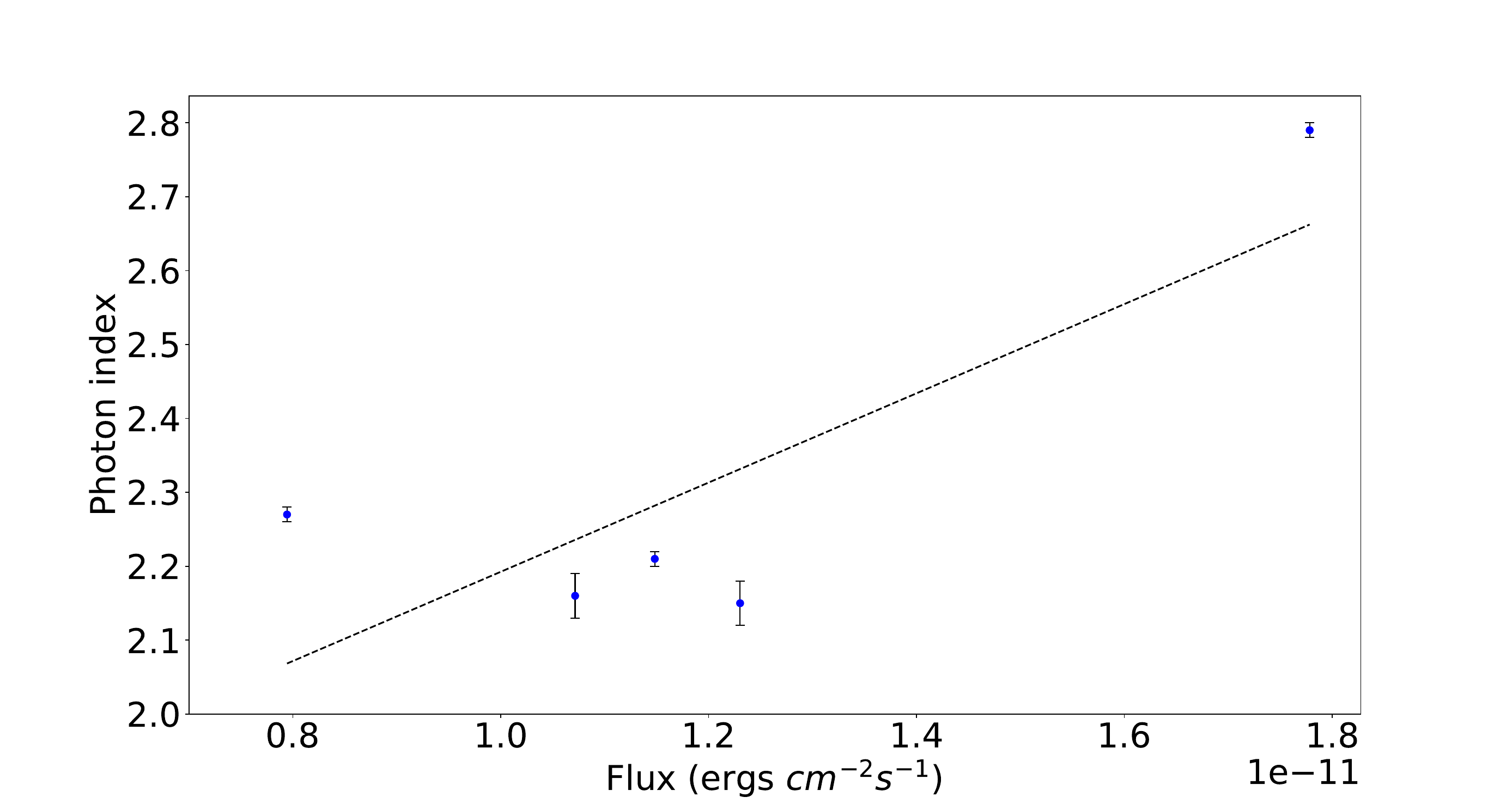}
    \caption{Photon index from logparabola model plotted against flux in the total energy band}
    \label{logPIvsflux}
\end{figure}

The spectral fitting of each observation with different spectral models is shown in table \ref{Spectral_fitting_nH_frozen_II}.
We notice the following spectral behavior in the \textit{XMM-Newton} observations of BL Lacertae:
\begin{itemize}
    \item In the first observation, 0501660201 (observed on 10-07-2007), the $\chi^{2}$ fitting is below the expected value of 1. The AICc value is smallest for powerlaw followed by logparabola model. These two models appear to fit the observation 0501660201 quite well compared to broken power-law and double power-law which also show a huge error against $\chi^{2}$ fitting. 
    The broken powerlaw model has the lowest AICc value compared to all the other models fitted to 0501660401 (observed on 08-01-2008) and for the four models fitted to 0501660301 (observed on 05-12-2007). 
    \item The variable observation 0504370401 (observed on  16-05-2008) is well-described by broken powerlaw models as explained by the lowest AICc value. The AICc values of models fitted to the variable observation 0891800501 (observed on 09-07-2021) shows double powerlaw model to be the best model describing the X-ray emission. The reduced $\Chi^{2}$ of the fitted broken power-law model suggests that it is also a good model. Thus, we pick broken powerlaw model as a good choice to evaluate the upturn energy in SED at which synchrotron hump gradually moves to inverse Compton hump.  
    \item Each of the observations were also well-fitted by logaparbola model, described by a negative curvature, with $\chi^{2}$ value within acceptable limit. The negative curvature obtained from this model pinpoints the existence of the upturn from synchrotron to inverse Compton emission in the SED of BL Lacertae in 0.3 - 10.0 keV energy range. The logparabola photon index given by alpha is higher in the flaring observation than in other observations giving an indication of the observation going in a soft state during the flare in 2021. The value of photon index for 0891800501 (date 09-07-2021) is above 2.5 while for all other observations it is below 2.5.
    To confirm the softer when brighter behaviour, the spectral parameter of the logparabola model was plotted against flux in figure \ref{logPIvsflux}. 
    Softer when brighter behavior is evident in BL Lacertae. An increased negative curvature ($\beta$ = -0.77) has been observed in the brightest observation which is consistent with the softer when brighter behavior and indicating a steep high-energy tail from inverse Compton contribution in BL Lacertae's X-ray spectrum. The modeling with double power-law model also helped identify the presence of both synchrotron and inverse Compton emission processes playing a role in the emission in X-ray energy range of 0.3-10.0 keV \citep{2025A&A...693A.299W}.
    \item All observations also showed good fit with blackbody model and have acceptable reduced $\chi^{2}$ values of closer to 1 (see table \ref{kt_tbabs0_II}).
    Addition of blackbody model was useful for getting the value of temperature kT of the disk which has been used to understand its change with varying flux. A positive trend of kT is observed with flux and was found highest for the observation 0891800501 observed to be flaring in 2021 (see figure \ref{Bbodyvsall_ph0}).
    \item The best fit models were obtained with the galactic absorption nH kept frozen at 2.7 $\times$ $10^{21} cm^{-2}$ which is in accordance with the previous work of \citet{2022MNRAS.509...52D} and takes into account the intervening molecular cloud in the line of sight of BL Lacertae. To check for any additional absorption or deviation from the frozen absorption value, the best fitted broken powerlaw model was used. The parameters of broken powerlaw model for each absorption were kept frozen to the best fit parameters as obtained in table \ref{Spectral_fitting_nH_frozen_II} and nH was left to vary. Galactic absorption nH was found to be 2.71 $\times$ $10^{21} cm^{-2}$ for the three observations -- 0501660201, 0501660301 and 0501660401. Galactic absorption nH was found 2.70  $\times$ $10^{21} cm^{-2}$ and 2.68 $\times$ $10^{21} cm^{-2}$ for observations 0504370401 (observed on 16-05-2008) and 0891800501 (observed on 09-07-2021) respectively. 
    \end{itemize}

\subsection{Soft and hard bands- variability and correlation}
\begin{figure*}[htbp]
    \centering
    \begin{tabular}{cc}
(a)
\includegraphics[width=65mm, height=60mm]{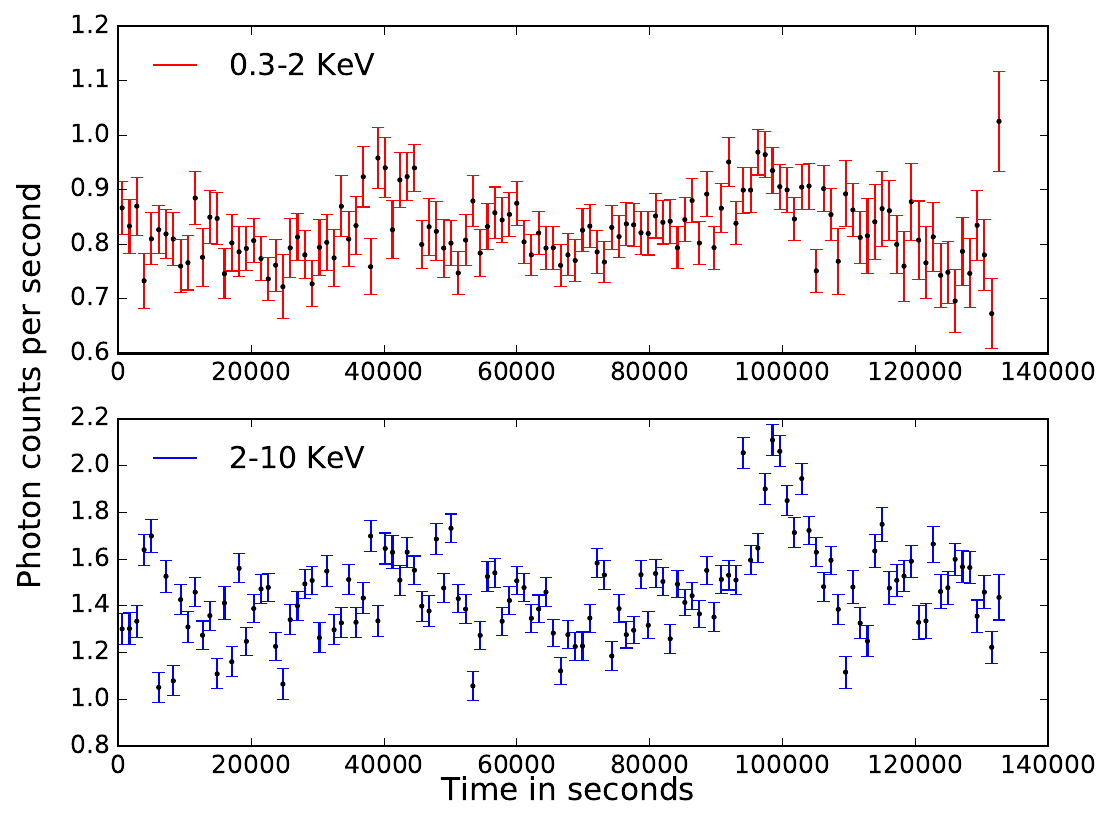} &
(b)
\includegraphics[width=65mm, height=60mm]{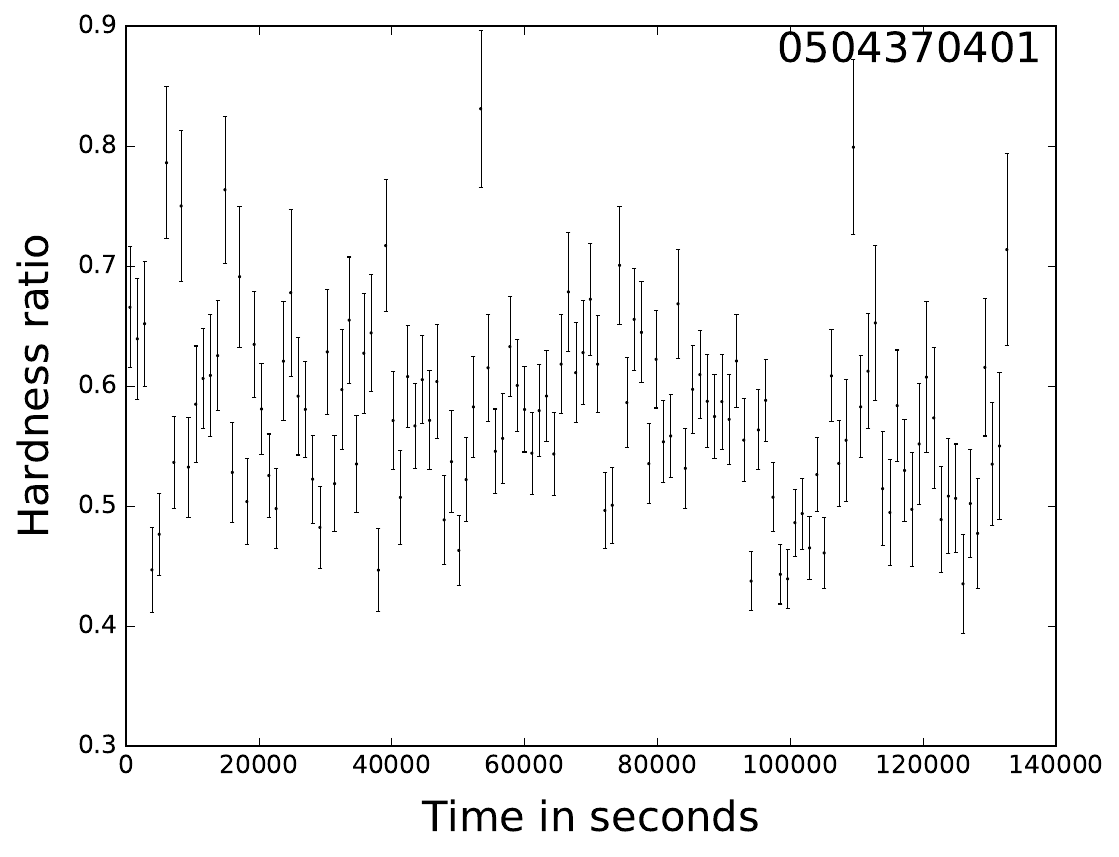} \\  
(c)
\includegraphics[scale=0.44]{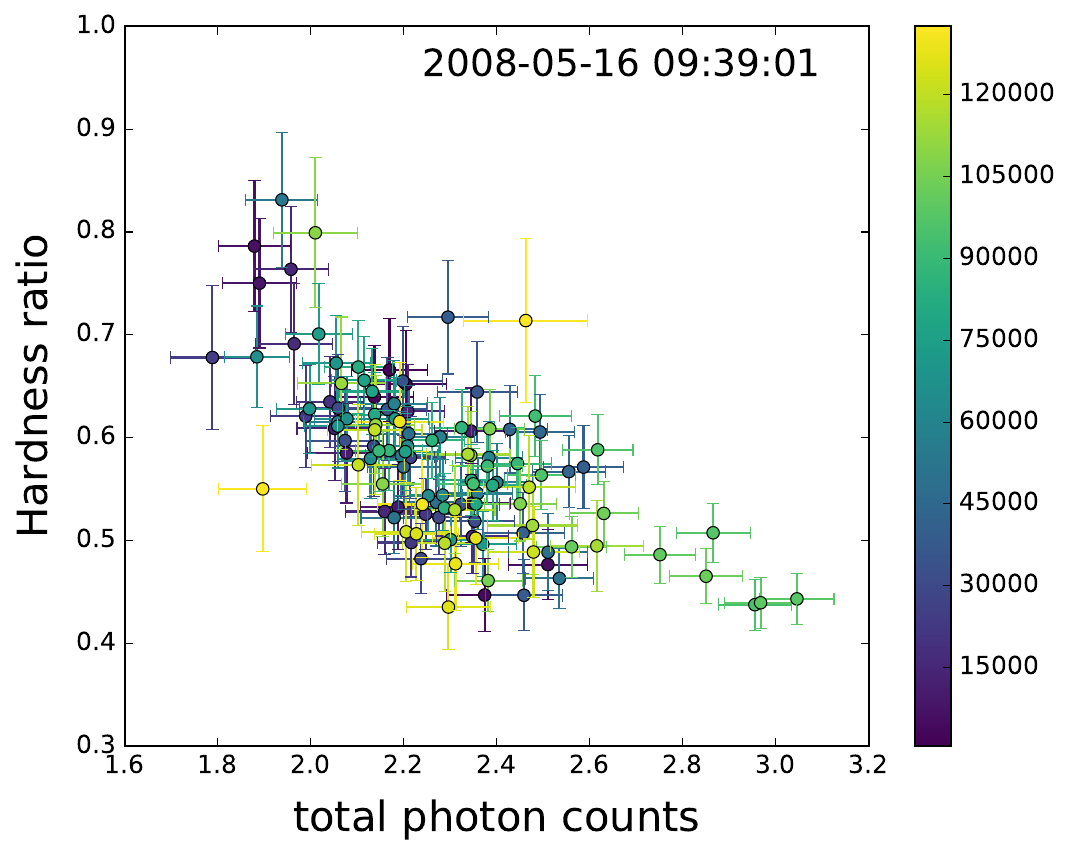} &
(d)
\includegraphics[width=65mm, height=60mm]{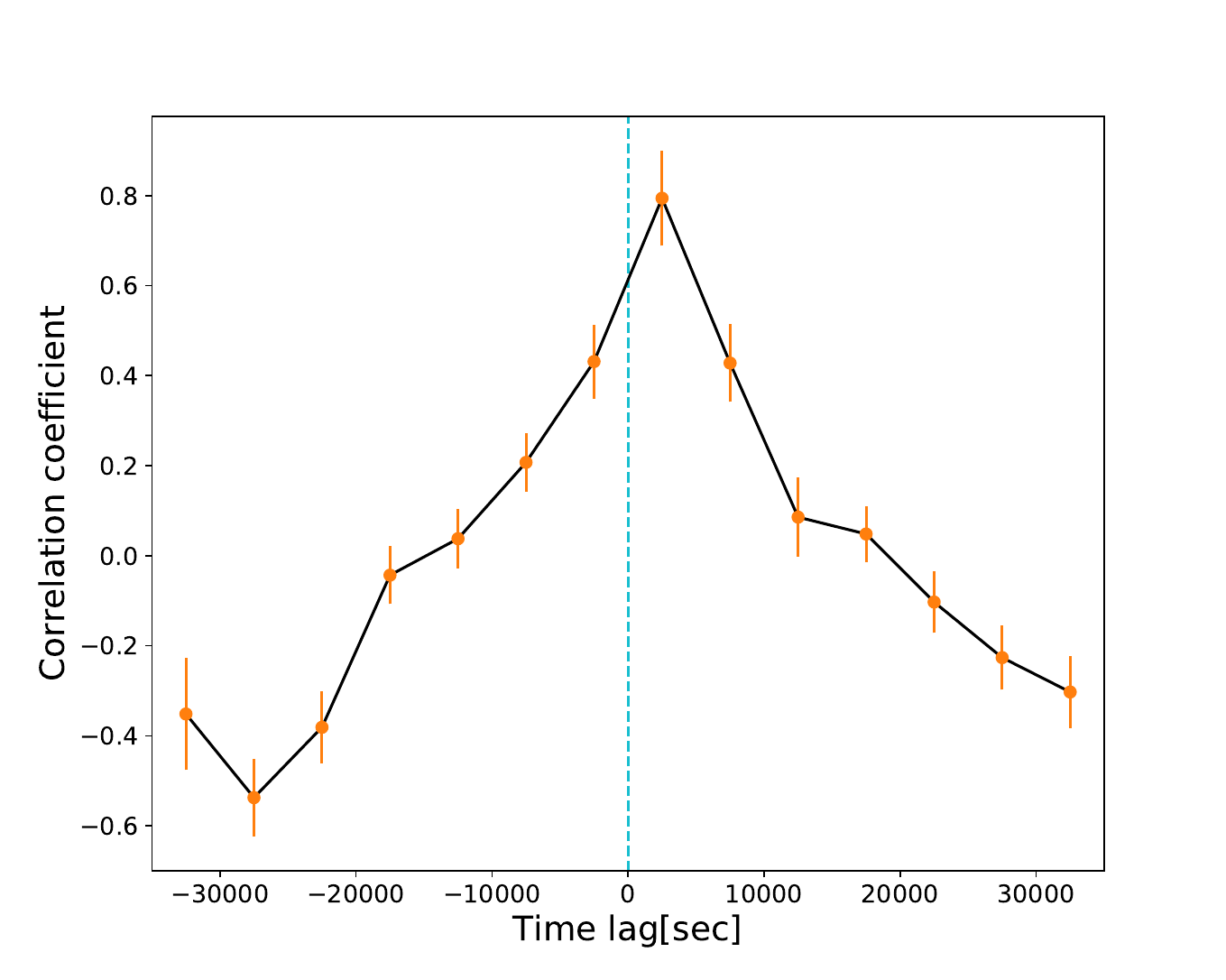} \\   
\end{tabular}
\caption{Observation 0504370401 (a) light curves in soft (0.3 - 2.0 keV) and hard band (2.0 - 10.0 keV) with 1100s time binned (b) hardness ratio against time (c) hardness ratio against intensity in which the colourbar represents the timescale in seconds from beginning of observation to end of it (d) discrete correlation function estimated between hard and soft energy range.
}
\label{HR_0504}
\end{figure*}

\begin{figure*}[htbp]
    \centering
    \begin{tabular}{cc}
(a)
\includegraphics[width=65mm, height=60mm]{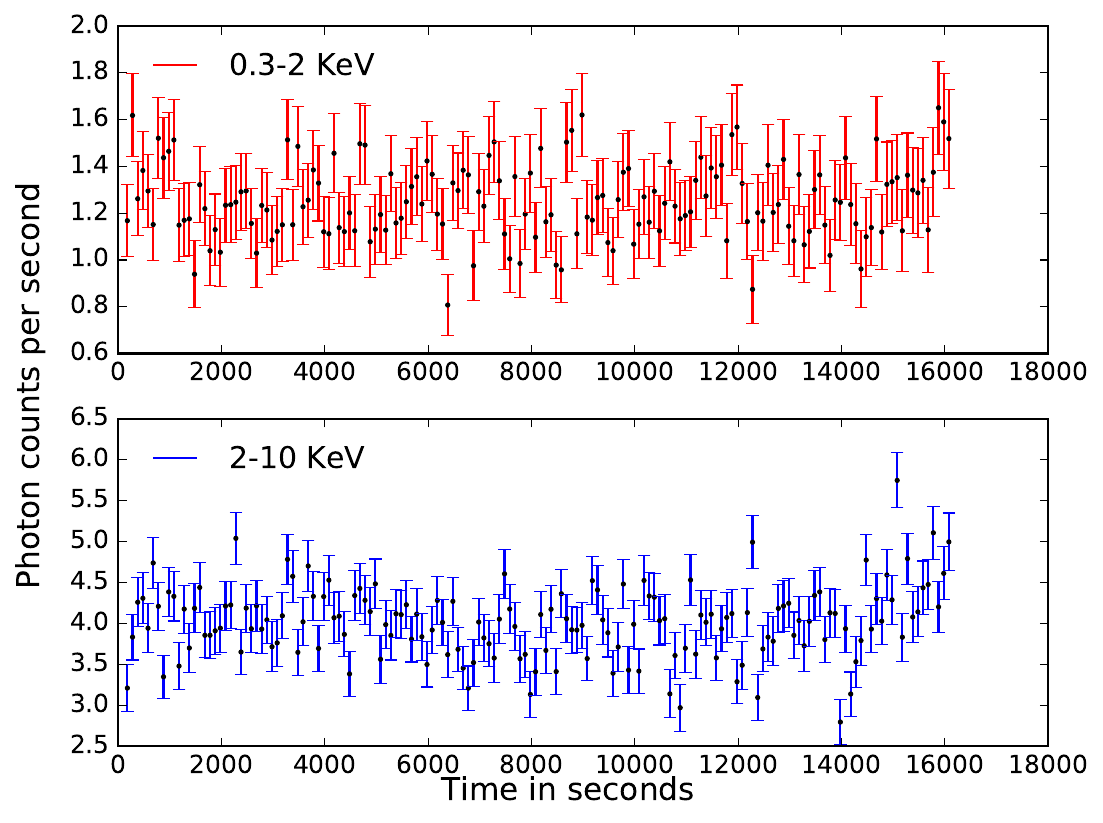} &
(b)
\includegraphics[width=65mm, height=60mm]{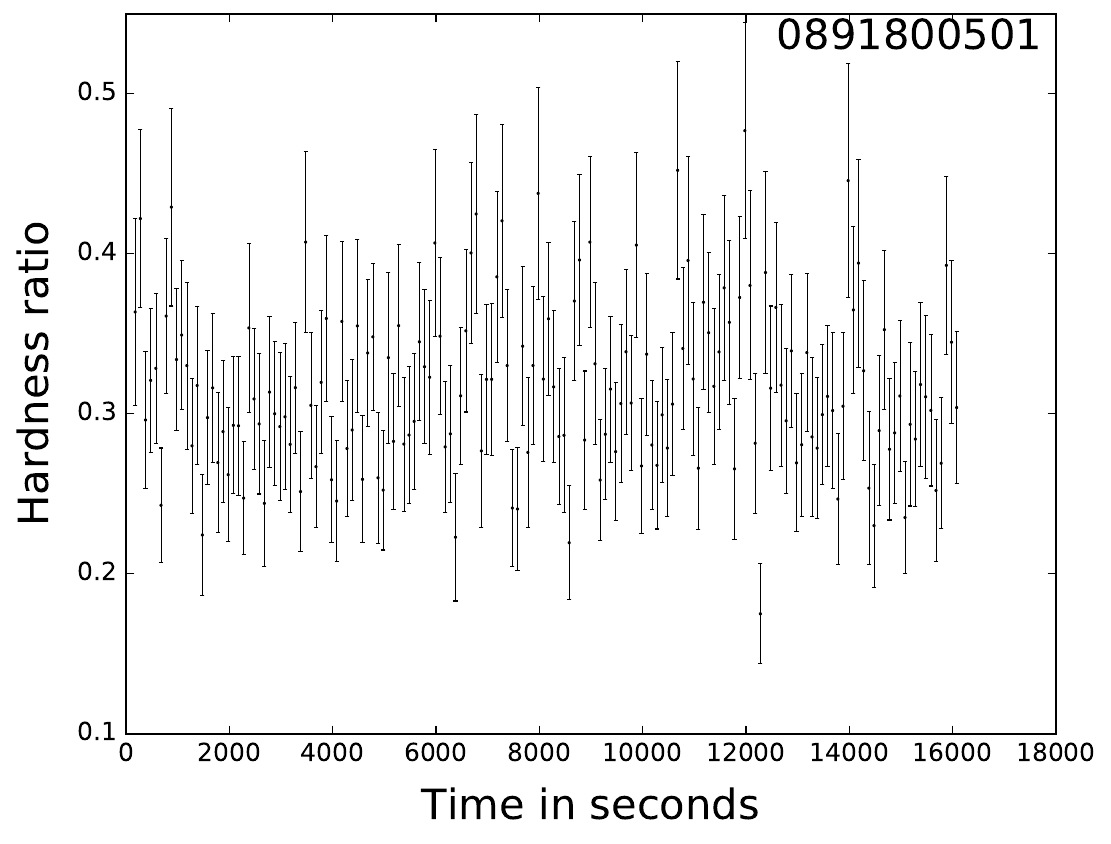} \\  
(c)
\includegraphics[width=65mm, height=60mm]{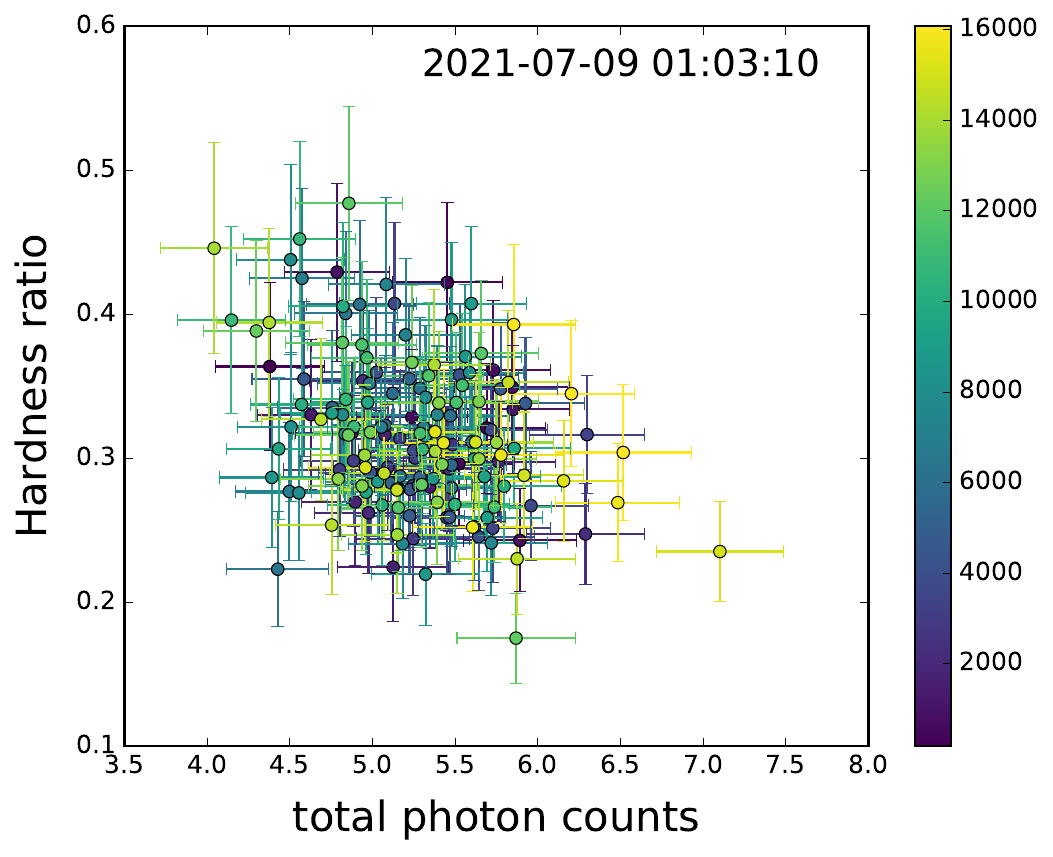} &
(d)
\includegraphics[width=65mm, height=60mm]{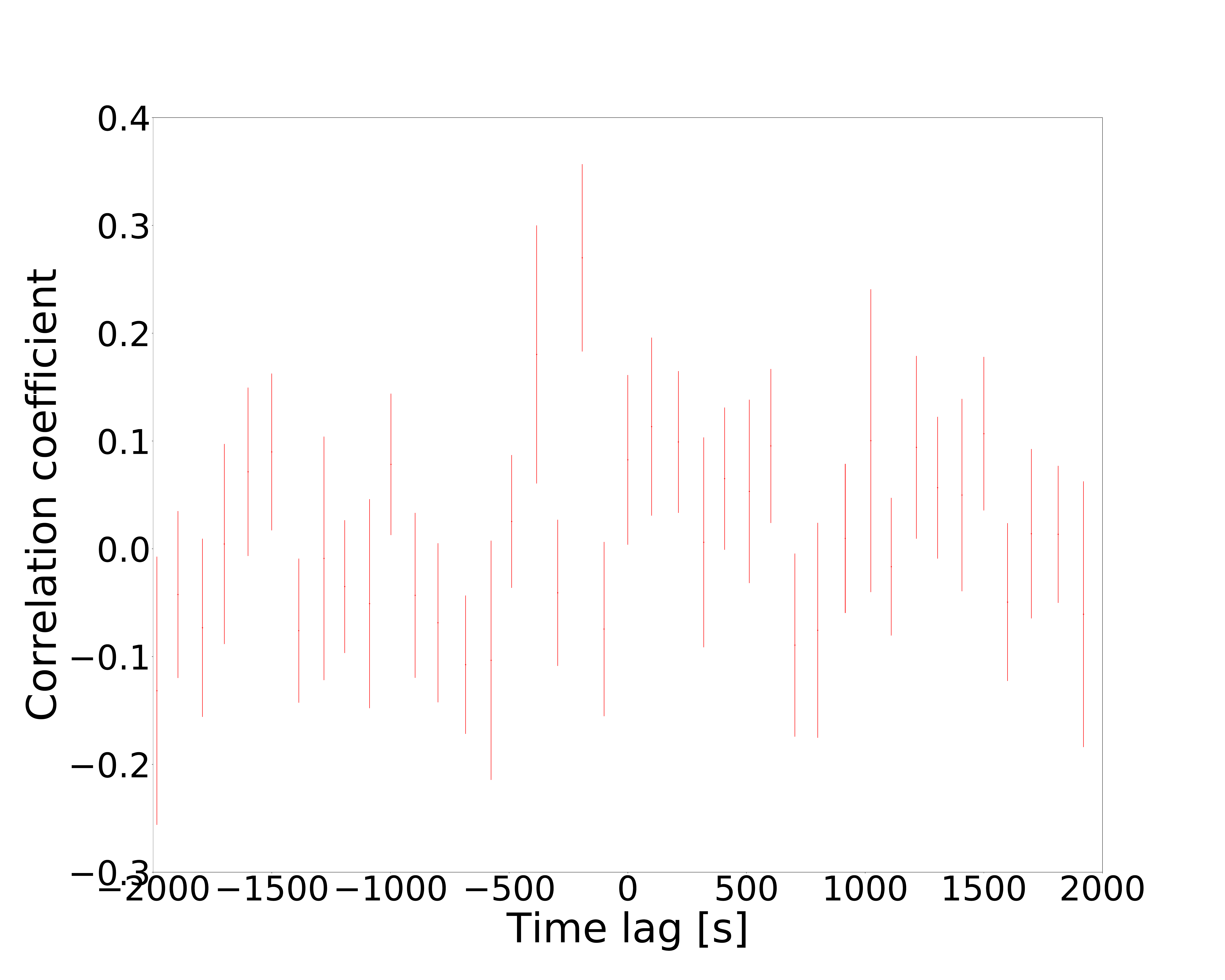} \\   
\end{tabular}
\caption{Observation 0891800501 (a) light curves in soft (0.3 - 2.0 keV) and hard band (2.0 - 10.0 keV) (b) hardness ratio against time (c) hardness ratio against intensity (d) discrete correlation function estimated between hard and soft energy range. 
}
\label{HR_08918}
\end{figure*}

The two variable observations were analysed for change of hardness ratio with time and flux. The correlation between the soft and hard bands was tested using DCF. Below are the results for the two variable observations: \\
\\
\textit{0504370401}, In this non-flaring variable observation (observed on 16-05-2008), there is no trend of hardness ratio observed with time (Figure \ref{HR_0504}(b)). The emission gets softer with increasing flux in this observation which is evident in Figure \ref{HR_0504}(c). 
The DCF peak reaches a significance of ~1.5$\sigma$ above the 1$\sigma$ confidence band but does not exceed the 2$\sigma$ or 3$\sigma$ thresholds derived from our Monte Carlo simulations. This indicates that the apparent correlation is not statistically significant and may be consistent with noise. \\
\\
\textit{0891800501} In this flaring variable observation (observation date 09-07-2021), softer when brighter behavior is observed in the hardness ratio versus flux plot, given in Figure \ref{HR_08918}(c). The light curves in the soft and hard band also do not show any obvious correlation except that after 14000 seconds of the start time of observation, the two light curves show a slight increase in flux count.  While this nature of softer when brighter is consistent with \textit{Swift}-XRT observations of BL Lacertae during its October 2020 flare \citep{2021MNRAS.507.5602P}, this is opposite to what was observed in the source during its flare in 2012 \citep{2022MNRAS.513.4645S}.  There is no trend in hardness ratio with respect to time. 
The DCF plot, Figure \ref{HR_08918}(d), is randomly distributed and did not show any strong correlation between the two wavebands.
DCF peak shows a marginal significance of ~1.3$\sigma$ above the 1$\sigma$ confidence band but does not exceed the 2$\sigma$ or 3$\sigma$ levels in Monte Carlo simulations performed, indicating that the correlation is not statistically significant. 

\subsection{Interval-wise spectral variability}
The two variable observations- 0504370401 (date 16-05-2008) and 0891800501 (date 09-07-2021) were split into shorter time intervals. The intervals of each observation are shown in their light curve in figure\ref{LC_var}. The observation 0504370401 is of longer time duration and its light curve is showing many variable states. Hence the observation was split into five different sections. Observation 0891800501 is a shorter observation with 19.6 ks. This observation was split into two sections. The first interval of this observation shows no visible variability in the light curve, whereas the second interval exhibits an increasing flux. Spectral fitting was performed separately for each interval. \\
From the previous section, we are aware of the concave structure of the x-ray spectra of BL Lacertae from the fitting of the logparabola to all the \textit{XMM-Newton} observations of BL Lacertae. Also, the broken power-law was a good fit for all observations. Thus, we proceeded with fitting each interval with the broken power-law model which gives the value of break energy ($E_{b}$).  $E_{b}$ parameter of the broken power-law model gives the energy at which the SED transitions from the first hump to the second hump. Thus, the part before the break energy is from the contribution of synchrotron emission, and after the break energy lies the inverse Compton component.  The spectral fitting of broken power-law fitted to the spectra of each section of observation is given in table
\ref{bknpl_sections_II}. Here it is visible that there is variability in spectral fitting in each section of the two variable observations. The break energy is changing. 
The flux for each section below and above break energy is given in Table \ref{synchrotron_contribution_II}.  
The fitted model to the observations are in online supplementary material. 
\\
Break energy from broken power-law is plotted against flux in the total energy band for each section of the two variable observations in figure \ref{Ebvsflux_ph0}.
In figure \ref{synchvsflux_ph0},  we plotted a variation of synchrotron contribution against flux and break energy obtained from broken power-law fitting against total flux respectively for each section of the variable observations. Synchrotron contribution is higher in the section with a higher flux value.
Both the plots (figures \ref{Ebvsflux_ph0} and \ref{synchvsflux_ph0}) show how the synchrotron contribution is high in observation with higher flux for this source. Thus, it can be said that synchrotron emission played a role in the X-ray flaring of BL Lacertae in 2021. 

\section{Discussion}
\begin{figure*}
    \centering
    \includegraphics[scale=0.4]{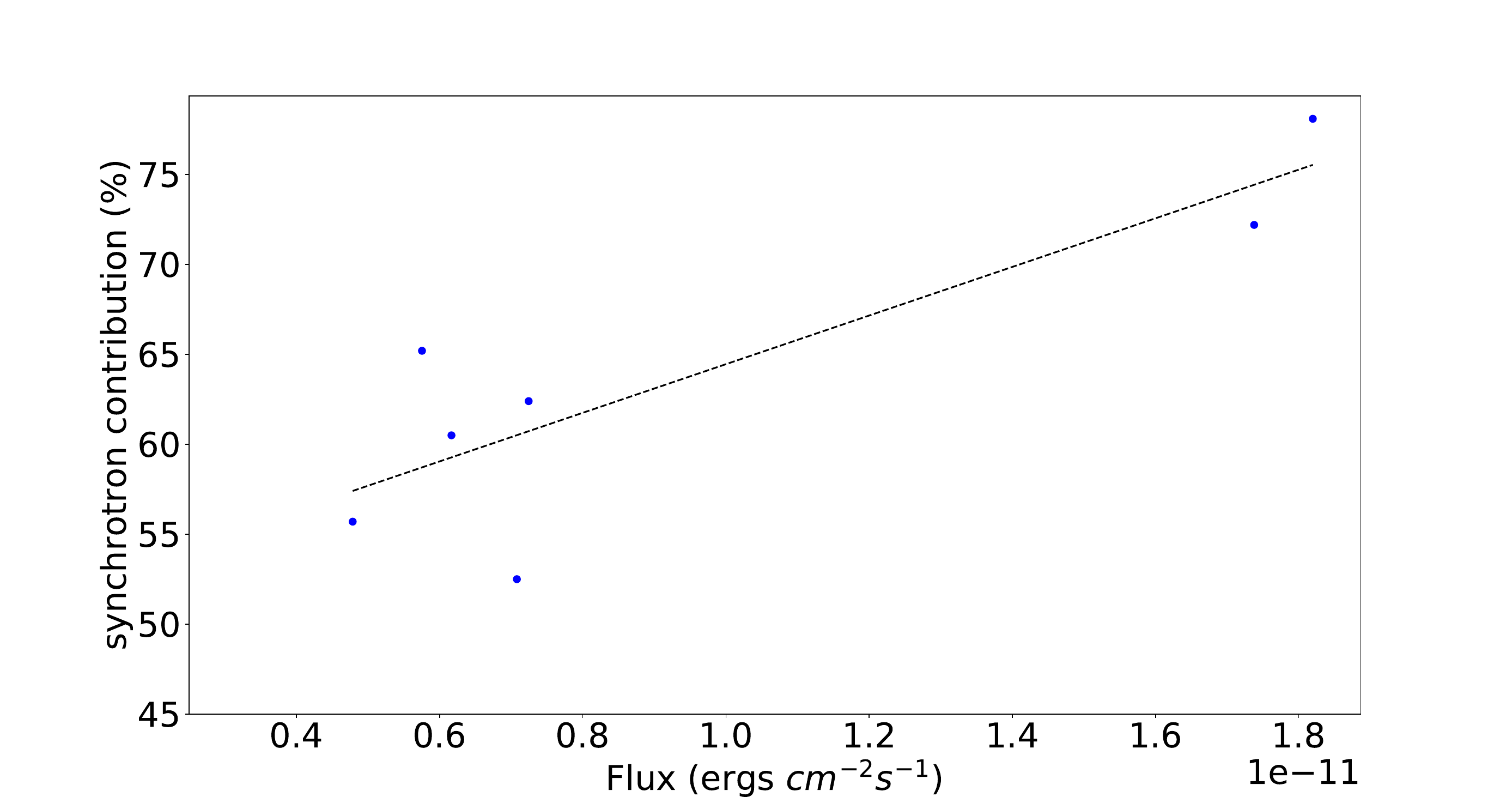}
    \caption{Contribution from synchrotron emission plotted against flux over the entire energy range for each section of the two variable observations}
    \label{synchvsflux_ph0}
\end{figure*}

\begin{figure*}
    \centering
    \includegraphics[scale=0.4]{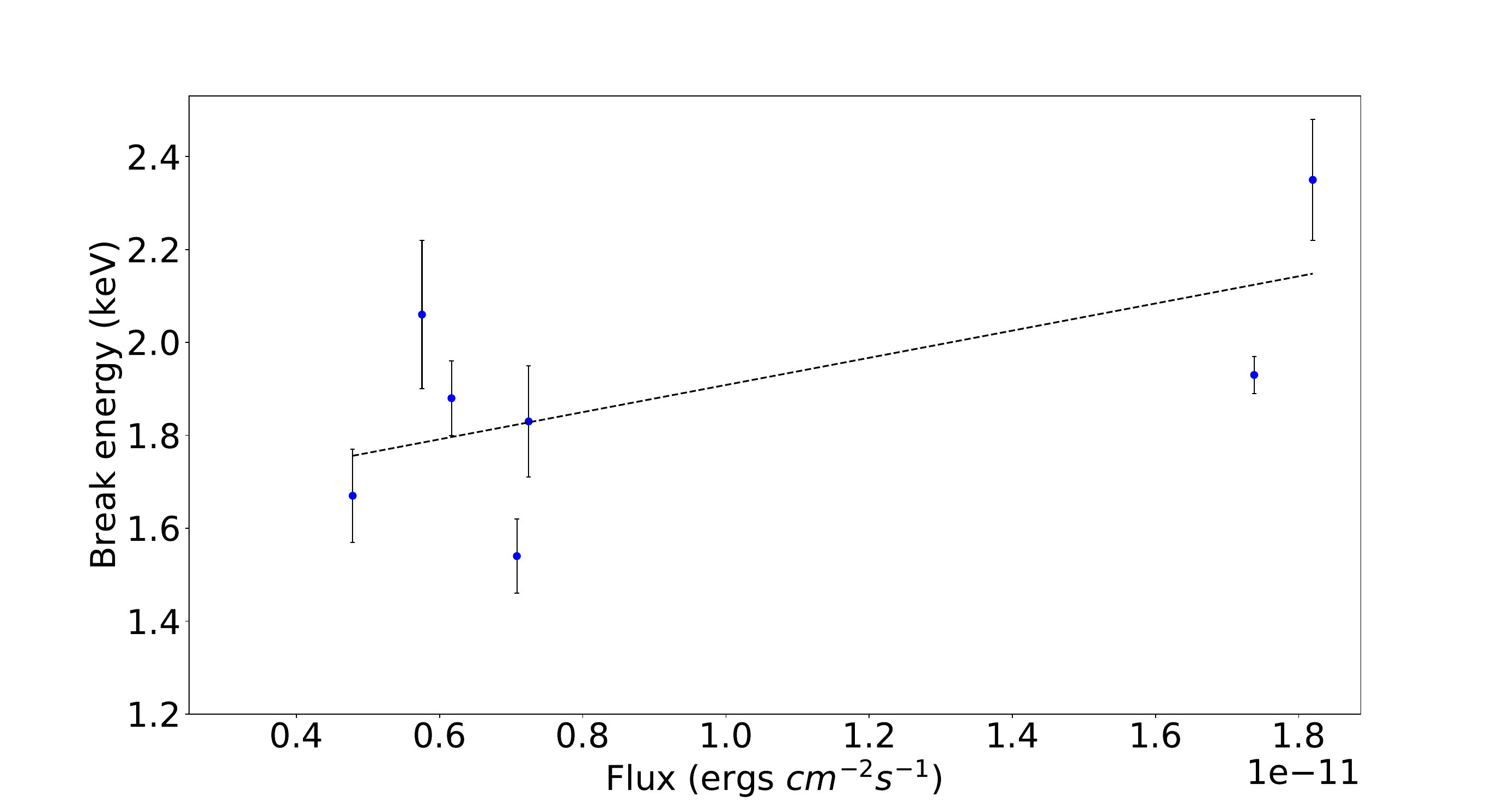}
    \caption{Break energy from broken power-law plotted against flux over the entire energy range for each section of the two variable observations}
    \label{Ebvsflux_ph0}
\end{figure*}

\begin{figure*}[htbp]
    \centering
    \begin{tabular}{c}
(a)  \\
\includegraphics[scale = 0.25]{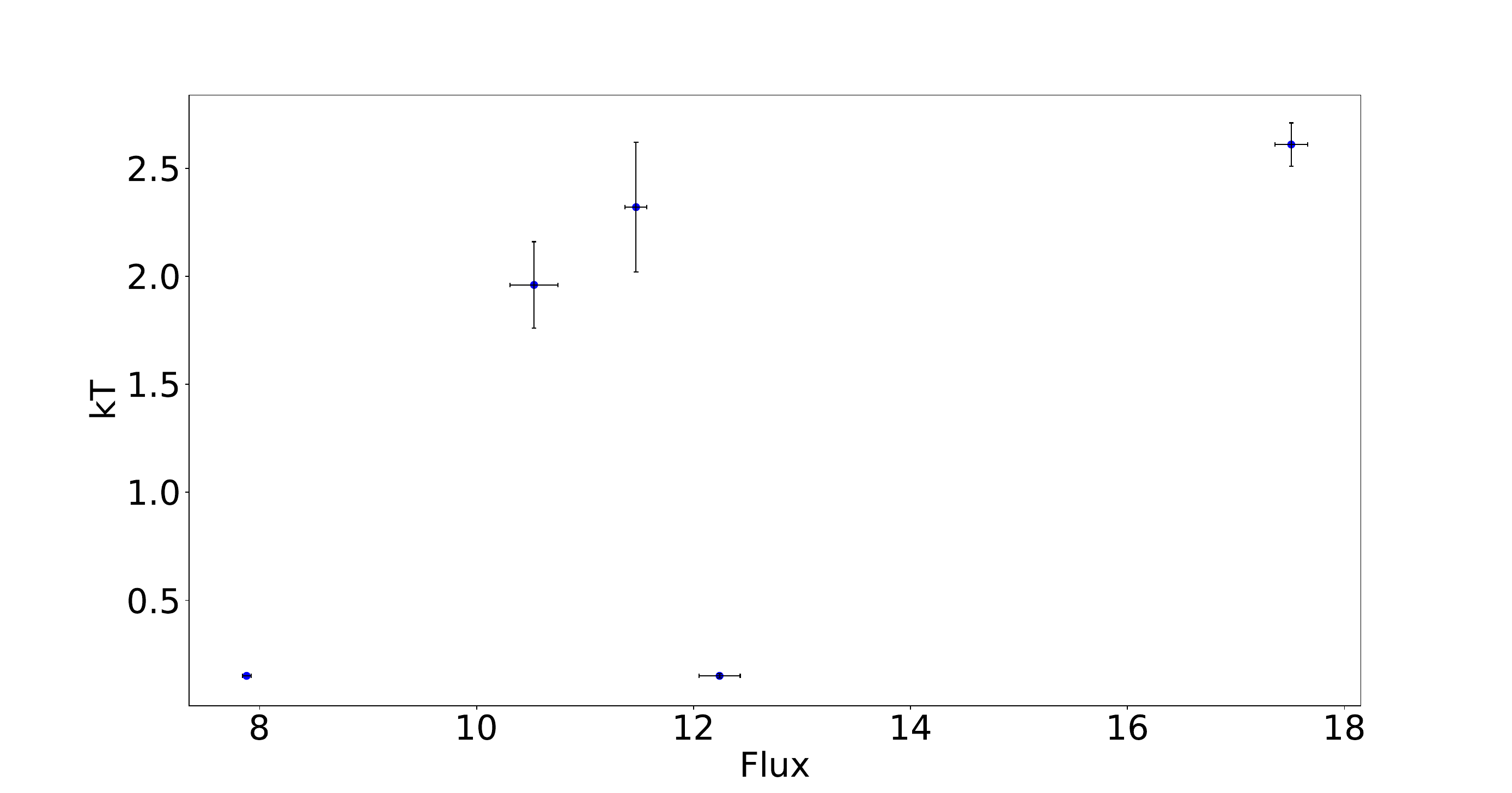} \\
(b) \\
\includegraphics[scale = 0.25]{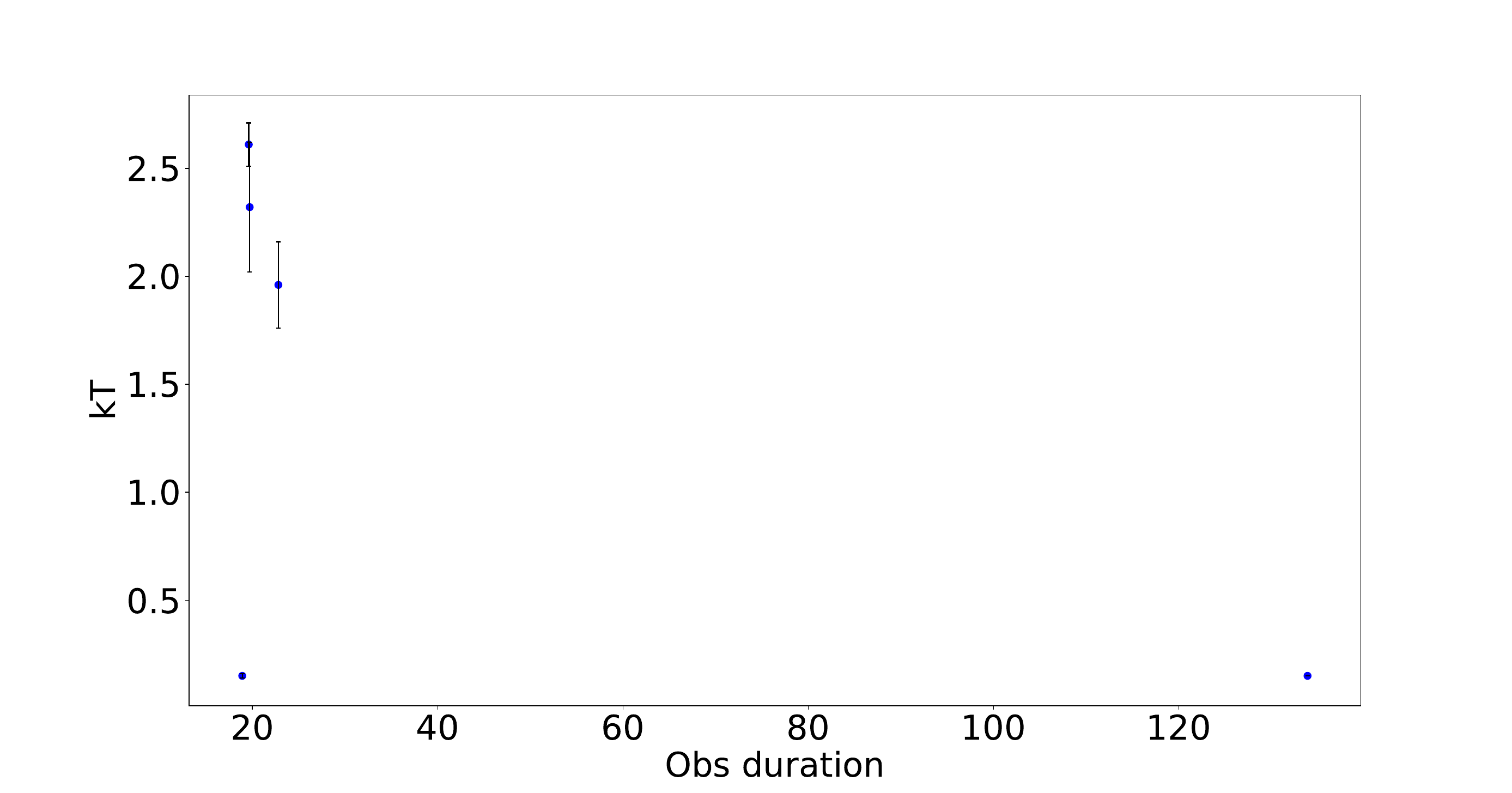} \\
(c) \\
\includegraphics[scale = 0.25]{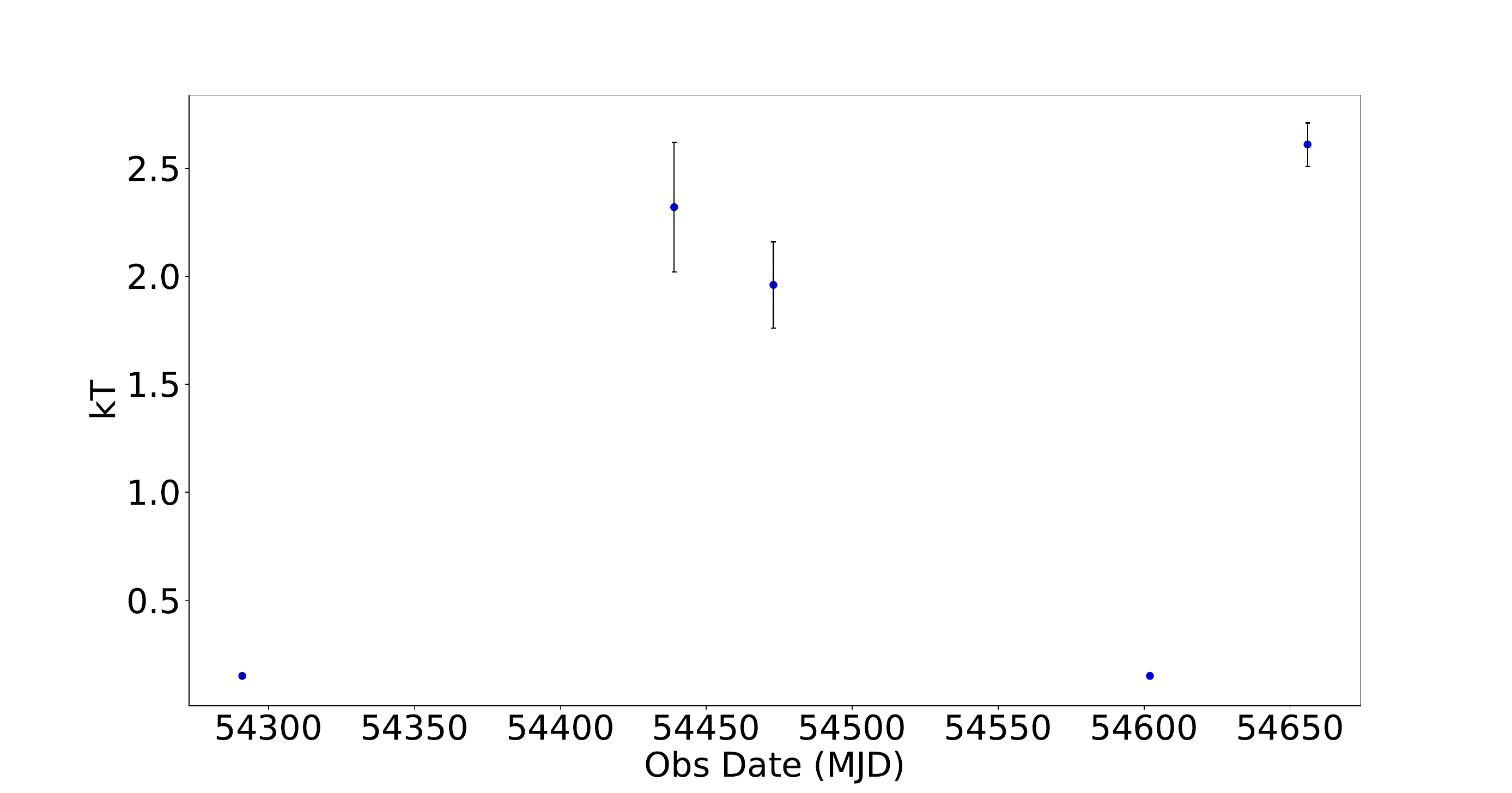} \\  
\end{tabular}
\caption{(a) Variation of blackbody temperature kT given in keV with respect to flux. (b) kT plotted against time duration of observation. (c) Variation of kT with the date of observation. }
\label{Bbodyvsall_ph0}
\end{figure*}

BL Lacertae exhibits complex behavior, displaying variability ranging from low to extremely high levels. In our set of observations, two out of five (0504370401, taken on 16 May 2008, and 0891800501, taken on 9 July 2008) show high variability, while the remaining three display low variability. One of the two variable observation has a very long exposure time. This appears in accordance with the previous finding of a moderate correlation between exposure time and fractional variability by \citet{2022ApJS..262....4N} in the blazar Markarian 421. \\
BL Lacertae has been observed to exhibit LBL characteristics \citep{2018A&A...620A.185N}, IBL characteristics \citep{2011ApJ...743..171A, 2016A&A...592A..22H}, and HBL characteristics \citep{2021MNRAS.507.5602P} at different times. 
The negative curvature of the log-parabola model and the soft-to-hard spectral break of the broken power-law model both indicate a concave spectrum, suggesting the presence of two distinct emission components or a spectral minimum in the 0.3-10.0 keV band. The broken power-law model shows a softer first photon index compared to the harder second one, with the break energy ranging from 1.81 keV to 2.58 keV across the observations. A spectral break in this range implies that the synchrotron peak frequency is located at or below these energies. This behavior is characteristic of BL Lacertae acting as an Intermediate-peaked BL Lac (IBL) object during these epochs.
\\
This source has been well studied in the past, with several works aiming to estimate the exact location of the synchrotron peak \citep{2009A&A...507..769R, 2018A&A...620A.185N, 2021MNRAS.507.5602P}. \citet{2009A&A...507..769R} compiled the broadband spectrum using simultaneous observations across the wavebands (from radio to X-ray) and showed that the synchrotron peak lies in the near-IR band. They also noted that the X-ray spectrum is highly variable, changing from extremely steep to extremely hard during different flux states. \\
The Galactic absorption obtained from the analysis is significantly higher than the survey value of galactic absorption obtained in HI4PI survey \citep{2016A&A...594A.116H}. The excess absorption suggests the presence of transient absorbing material - a molecular cloud along the line of sight to the emission region. This molecular cloud has been extensively studied earlier using ${}^{12}$CO observations by \citet{1991AJ....101.2147B} and \citet{1998A&A...339..561L}. This value is in compliance with the work of \citet{2022MNRAS.509...52D} and much lower than \citet{2009A&A...507..769R}. \\
The addition of a blackbody component was statistically required to model a prominent soft X-ray excess in the spectrum. A similar feature was noted by \citet{2009A&A...507..769R}, who argued that its variability pointed to a dynamic origin. Since the non-thermal jet emission was bright during the observation 0891800501 (observed on 09-07-2021), this soft excess likely represents an additional emission component rather than the unveiling of a faint, underlying disk.  Future multi-wavelength campaigns incorporating optical/UV monitoring and X-ray polarimetry will be crucial to distinguish between an accretion-based or jet-based origin for this excess. \\
In this study, we found that as the source transitions from a low-flux state to a high-flux state, the photon index changes from 2.2 to 2.9, indicating softer-when-brighter behaviour as shown in Figure 3. Similar results were also reported by \citet{2021MNRAS.507.5602P} using \textit{Swift}-XRT spectra. While BL Lacertae has previously displayed the “harder-when-brighter” behavior typical of blazars, it showed a “softer-when-brighter” trend in the two variable observations examined here, consistent with the findings of \citet{2021MNRAS.507.5602P} during the 2020–2021 flare of BL Lacertae.
\\
The contribution of synchrotron emission in the 0.2–10.0 keV energy range appears to increase with flux (figure \ref{synchvsflux_ph0}). Additionally, the correlation between the soft and hard bands showed no significant correlation in the variable observations. The contribution of the synchrotron component in the 0.3–10.0 keV range increased during the flaring state, indicating that enhanced synchrotron emission is one of the drivers of the source's flaring activity. Since the soft and hard sub-energy bands show no correlation, the synchrotron-driven flaring may be attributed to factors unrelated to single emission zone such as interaction with external medium, geometric effects such as precession, or multiple emission regions.
\\
BL Lacertae was observed to soften during the flaring state. This could be explained by the extension of synchrotron emission to higher energies, with the inverse Compton component remaining constant. This behavior was also verified by the correlation of break energy with flux. A similar observation was made for S5 0716+71 by \citet{1999A&A...351...59G}, \citet{2006A&A...457..133F}, and \citet{2003A&A...400..477T}. 
\\
IBLs are complex sources exhibiting variable spectral behaviors over time. Synchrotron emission appears to be a dominant mechanism driving the emission during the flaring in 0.3 - 10.0 keV in both BL Lacertae and S5 0716+71 \citep{2006A&A...457..133F}.
The concave structure observed in the log-parabola model fitted to BL Lacertae was also detected in the intermediate BL Lac object S5 0716+71 \citep{2006A&A...457..133F}. This feature was attributed to the transition from the tail of the synchrotron bump into the lower-energy region of the inverse Compton peak is similar to the log-parabola model fitted to BL Lacertae. The IBL S5 0716+71 also exhibited strong variability on timescales of hours. 
BL Lacertae exhibits similar behavior to other IBLs, such as S5 0716+71 and ON 231. However, it is also important to recognise its demonstration of both LBL and HBL behavior, as highlighted by previous studies. Only a detailed multiwavelength study of the SED of BL Lacertae conducted over an extended period of at least a decade (given its flaring pattern every 10 years) can fully explain its complex behavior and determine the exact classification of this blazar.
\\

\onecolumn
\begin{table}
\centering
\begin{threeparttable}
\small
    \renewcommand{\arraystretch}{1.5}
    \begin{tabular}{|l|l|l|l|l|l|l|l|l|}
    \hline
    Ob. ID. & model & $\Gamma_{1}$ or $\alpha$ & $\Gamma_{2}$ or $\beta$ & $E_{break}$ & $\chi^{2}$/dof & $Flux_{0.3 - 10.0}$ & AICc \\ \hline
        0501660201 & PL &  $2.14\pm0.01$ & - & -  & $0.91/121$ & $12.22 \pm 0.01$ & 114.21 \\ \hline
        - & LP   &$2.15\pm0.03$  & $-0.03 \pm 0.05$ & - & $0.91/120$ & $12.29 \pm 0.19$ & 115.40 \\ \hline
        - & BP & $2.16 \pm 0.04$ & $1.55\pm1.08$ & $2.11\pm 0.03$ &  0.92/119 & $12.30 \pm 0.50 $ & 117.02 \\ \hline
        - & DP  & $2.14 \pm 0.02$ & $-2.06^{ *}$ & - &  0.91/119 & $12.41 \pm 0.27$ & 114.56 \\ \hline
        0501660301 & PL  & $2.14\pm0.01$ & - & - &  1.25/135 & $11.18 \pm 0.01$ & 172.84 \\ \hline
        - & LP  & $2.21\pm0.01$ & $-0.14 \pm 0.02$ & - &  1.08/134 & $11.46 \pm 0.10$ & 150.90\\ \hline
        - & BP  & $2.23 \pm 0.02$ & $2.03 \pm 0.02$ & $1.86\pm0.25$ & 1.04/133 & $ 11.46 \pm 0.12 $ & 146.62 \\ \hline
        - & DP  & $2.49 \pm 0.05$ & $0.72\pm0.69$ & - & 1.05/133 & $11.53 \pm 0.01$ & 147.95 \\ \hline
        0501660401 & PL  & $2.09\pm0.01$ & - & - &  1.06/125 & $10.30 \pm 0.15$ & 136.60 \\ \hline
        - & LP  & $2.16\pm 0.03$ & $-0.12 \pm 0.04$ & -  & 1.02/124 & $10.57 \pm 0.22$ & 132.68 \\ \hline
        - & BP  & $2.16 \pm 0.03$ & $1.94\pm0.07$ & $2.58 \pm 0.65$ &  1.01/123 & $10.61 \pm 0.14$ & 132.56  \\ \hline
        - & DP  & $2.19 \pm 0.01$ & $0.80 \pm 1.44$ & - & 1.02/123 & $10.71 \pm 0.01$ & 133.79 \\ \hline
        0504370401 & PL  & $2.07\pm0.01$ & - & - & 7.47/172 & $7.76 \pm 0.04$ & 1288.91 \\ \hline
        - & LP  & $2.27\pm 0.01$ & $-0.41 \pm 0.01$ & - & 2.57/171 & $7.93 \pm 0.04$ & 445.61 \\ \hline
        - & BP  & $2.30 \pm 0.01$ & $1.79\pm 0.01$ & $1.81\pm0.04$ &  1.92/170 & $7.87 \pm 0.41$ & 334.64 \\ \hline
        - & DP  & $2.25 \pm 0.01$ & $0.72\pm0.05$ & - & 2.50/170 & $8.02 \pm0.05$ & 433.24 \\ \hline
        0891800501 & PL  & $2.46\pm0.01$ & - & - &  8.17/153 & $14.84 \pm 0.10$ & 1254.09 \\ \hline
        - & LP  & $2.79 \pm 0.01$ & $-0.77\pm0.02$ & - &  1.77/152 & $17.63 \pm 0.16$ & 275.20 \\ \hline
        - & BP  & $2.82 \pm 0.01$ & $1.82\pm 0.02$ & $1.98\pm0.04$ &   1.46/151 & $17.40 \pm 0.13$ & 228.73 \\ \hline
        - & DP  & $1.12 \pm 0.08$ & $3.06\pm0.04$ & - &  1.43/151 & $17.76 \pm 0.17$ & 224.20 \\ \hline
    \end{tabular}
    \caption{Spectral fitting of each observation. Each observation is corrected for galactic absorption model Tbabs of value 2.7 $\times$ $10^{21} cm^{-2}$. Here PL stands for power-law, LP for logparabola, BP for broken power-law and DP for double power-law. $\Gamma_{1}$ is the photon index of PL and first PL, and the first photon index of BP. In the case of LP, this becomes $\alpha$. $\Gamma_{2}$ is the second photon index of BP, and photon index of second PL in DP. In the case of LP, there is $\beta$ which gives the curvature of the model.  $E_{break}$ is the breaking energy of the broken power-law. Flux is in the unit of 10$^{-12} ergs cm^{-2}$ $s^{-1}$ in the energy range of 0.3 to 10.0 keV. The values where the error bars are high are mentioned as higherr in the table.}
    \label{Spectral_fitting_nH_frozen_II}
\begin{tablenotes}
\small
\item ${}^*$ $\Gamma_2$ was fixed at $-2.06$ because it was not constrained by the fit for observation 0501660201.
\end{tablenotes}
\end{threeparttable}
\end{table}
    
\begin{table}
\small
    \centering
    \renewcommand{\arraystretch}{1.5}
    \begin{tabular}{|c|c|c|c|c|c|c|}
    \hline
    Ob. ID  & Bbody  & logpar & logpar  & $\chi^2$/dof & $Flux_{0.3-10.0}$ & AICc\\
    & kT  & $\alpha$ & $\beta$  & & & \\
    \hline
      $0501660201$   &  $0.15 \pm 0.01$  &  $1.91 \pm 0.14$ & $0.18 \pm 0.14$ &  $0.88/118$  & $12.2 \pm 0.19$ & 119.07\\
       $0501660301$   & $2.32 \pm 0.29$ & $2.23 \pm 0.02$ & $0.15 \pm 0.11$ &  $1.03/132$ & $11.47\pm0.10$ & 146.42 \\
      $0501660401$    & $1.96 \pm 0.25$ & $2.19 \pm 0.04$ & $0.30 \pm 0.21$ &  $1.00/122$ & $10.53\pm0.22$ & 132.50 \\
       $0504370401$   & $0.15 \pm 0.01$ & $1.84 \pm 0.04$ & $-0.04 \pm 0.04$ &   $1.28/169$ & $7.88\pm0.04$ & 226.68 \\
      $0891800501$    & $2.16 \pm 0.11$ & $2.84 \pm 0.01$ & $-0.17 \pm 0.09$ &   $1.48/150$ & $17.51\pm0.15$ & 232.40 \\ 
      \hline
    \end{tabular}
    \caption{Spectral fitting of the five observations with blackbody model along with Galactic absorption Tbabs fixed at 2.7 $\times$ $10^{21} cm^{-2}$ and logparabola model. For a good fitting, a power-law model had to be added to the two variable observations: 0504370401 and 0891800501.}
    \label{kt_tbabs0_II}
\end{table}

\begin{table*}[!htbp]
\small
\caption{Spectral fitting of different sections of variable light curves with a broken power-law model with Galactic absorption Tbabs fixed at 2.7 $\times$ $10^{21} cm^{-2}$. Obs ID is observation ID. $\Gamma_1$ and $\Gamma_2$ are the two-photon index of broken power-law. Break energy ($E_b$) is in the unit of keV. FLux is in the unit of $10^{-12}$ ergs $cm^{-2}$ $s^{-1}$. $\chi^{2}$ is the reduced $\chi^{2}$ value with dof as degrees of freedom.}
\centering
\label{bknpl_sections_II}
\begin{tabular}{|c|c|c|c|c|c|c|}\hline 
$Obs ID$ & $section$ & $\Gamma_1$ & $E_b$ & $\Gamma_2$ & $flux_{0.3 - 10.0  keV}$ &  $\chi^{2}$ $(dof)$ \\
\hline
$0504370401$ & $I$ & $2.30\pm0.03$ & $1.67\pm0.10$ & $ 1.84\pm0.03 $ & $4.68 \pm 0.07$ & $1.27 (158)$ \\
non-  & $II$ & $2.29\pm0.03$ & $2.06\pm0.16$ & $1.75\pm0.05$ & $5.66 \pm 0.11$ & $1.16 (149)$ \\
flaring & $III$ & $2.28\pm0.02$ & $1.88\pm0.08$ & $1.76\pm0.02$ & $6.06\pm0.06$ & $1.18 (153)$ \\
& $IV$ & $2.36\pm0.03$ & $1.83\pm0.12$ & $1.86\pm0.03$ & $7.20 \pm 0.09$ & $1.38 (132)$ \\
& $V$ & $2.36\pm0.03$ & $1.54 \pm 0.08$ & $1.85\pm0.03$ & $6.96\pm 0.13$ & $1.12 (161)$ \\
\hline
$0891800501$ & $I$ & $2.83\pm0.01$ &  $1.93\pm0.04$ & $1.84\pm0.02$ & $17.23 \pm 0.14$ & $1.36 (148)$ \\
flaring & $II$ & $2.81 \pm 0.03$ & $2.35\pm0.13$ & $1.62\pm0.09$ & $18.09 \pm 0.39$ & $1.27 (106)$ \\
\hline
\end{tabular}
\end{table*}

\begin{table*}[htbp]
\small
\caption{Synchrotron contribution for each interval from the broken power-law model with frozen Tbabs}
\label{synchrotron_contribution_II}
\centering
\begin{tabular}{|c|c|c|c|c|c|}\hline 
Obs ID & section & break E & $flux_L$  & $flux_H$ & $\%$ synch \\  
 & & $E_b$ keV & $(0.3 - E_{b})$ keV & $(E_{b} - 10.0)$ keV & \\\hline
$0504370401$ & $I$ & 1.67 & 0.00073 & 0.00058 & 55.7 \\
-  & $II$ & 2.06 & 0.00103 & 0.00055 & 65.2 \\
 - & $III$ & 1.88 & 0.00101 & 0.00066 & 60.5 \\
 - & $IV$ & 1.83 & 0.00131 & 0.00079 & 62.4 \\
 - & $V$ & 1.54 & 0.00103 & 0.00093 & 52.5 \\
\hline
$0891800501$ & $I$ & 1.93 & 0.00424 & 0.00163  & 72.2 \\
- & $II$ & 2.35 & 0.00475 & 0.00133 & 78.1 \\
\hline
\end{tabular}
\end{table*}

\twocolumn

\begin{acknowledgement}
The project is co-financed by the Polish National Agency for Academic Exchange.
\end{acknowledgement}

\section*{References}
\bibliographystyle{aasjournal}   
\bibliography{example}
\onecolumn
\appendix       
\centering
\textbf{Appendix}
\begin{figure*}[!htbp]
    \centering
    \begin{tabular}{c}
\includegraphics[scale=0.45, angle = 0]{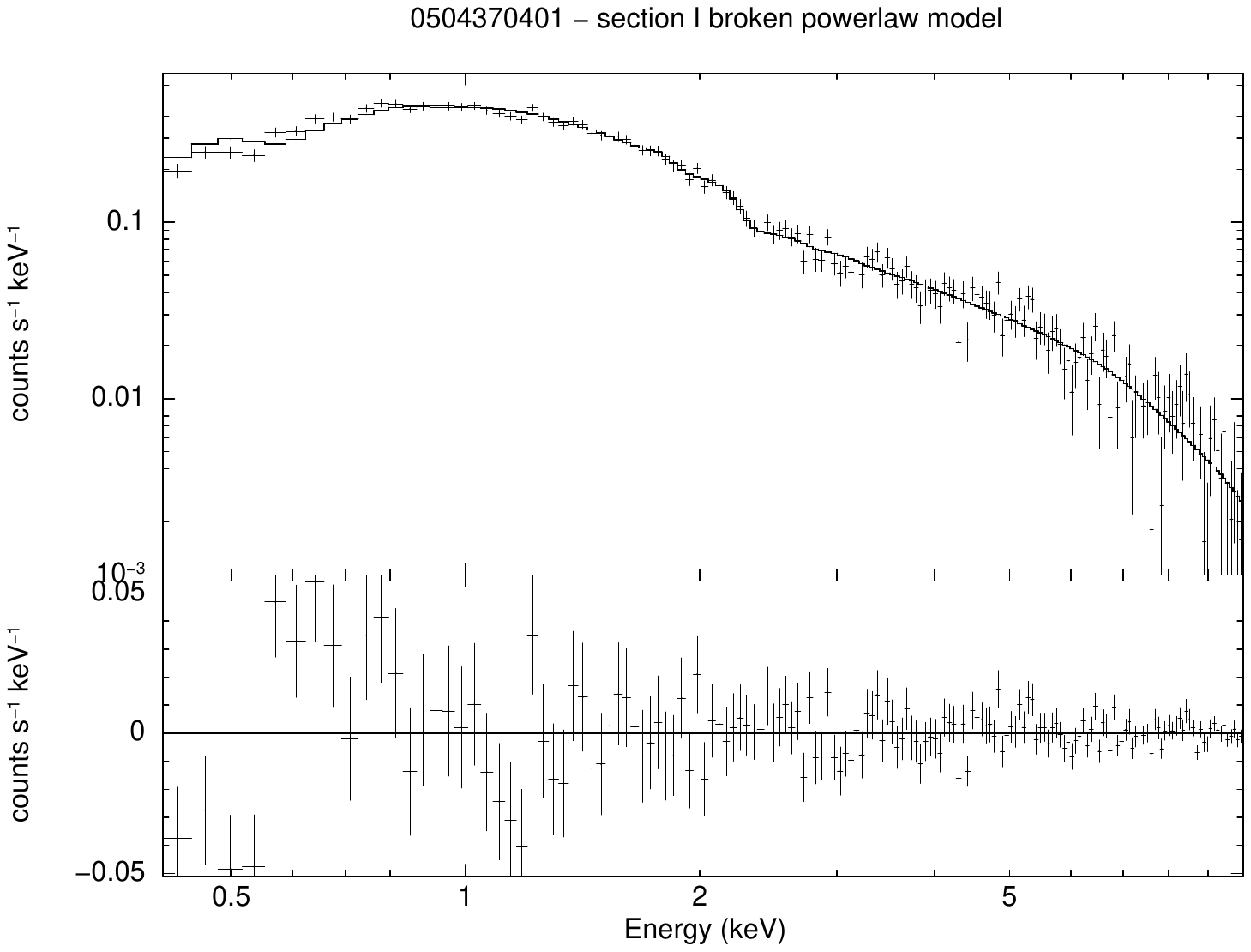} \\
\includegraphics[scale=0.45, angle = 0]{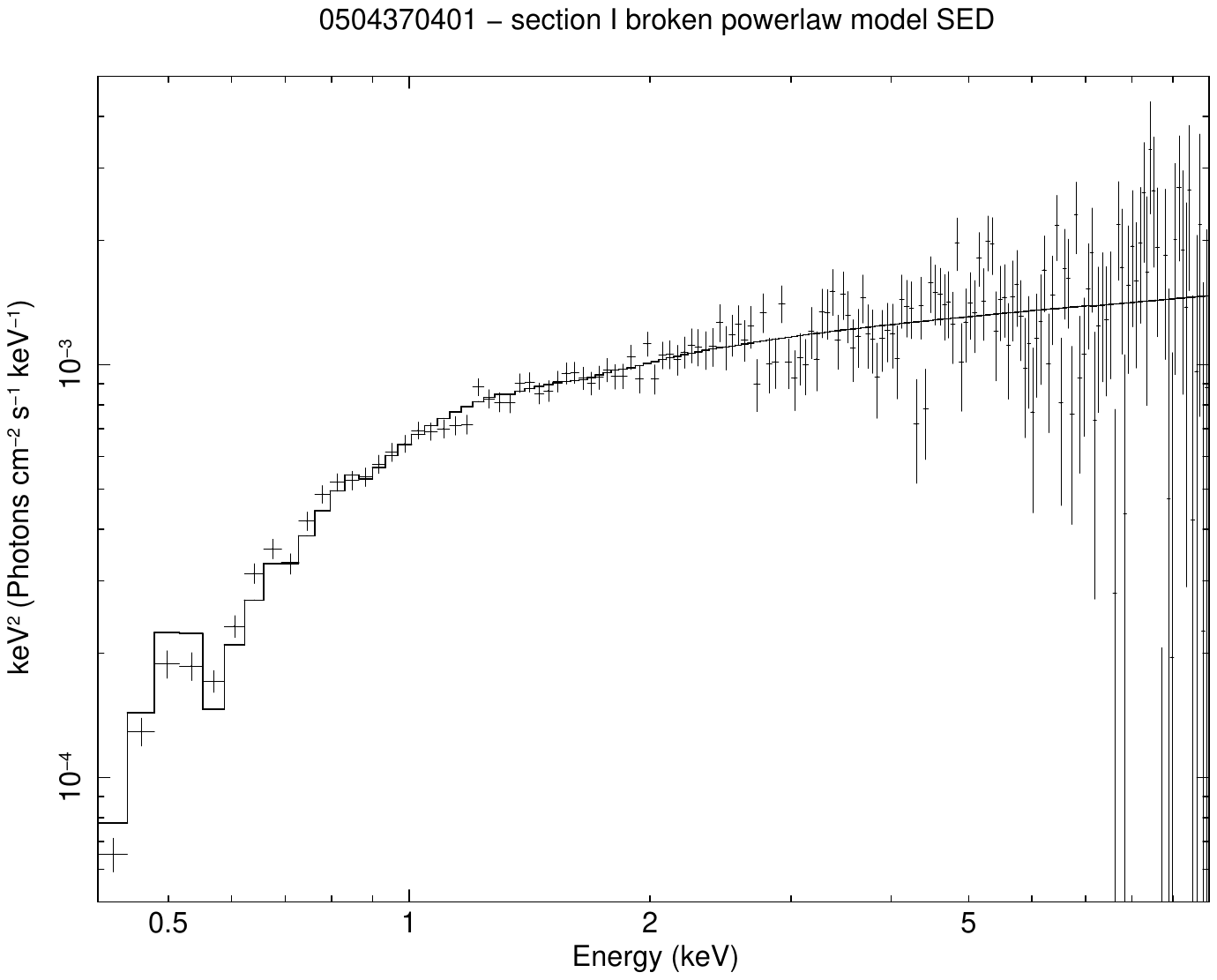} \\
\\
\end{tabular}
\caption{Top: Spectra fitted with broken power-law model along with the residuals. Bottom: SED plot for section I of observation 0504370401}
\label{050437spec_secI}
\end{figure*} 

\begin{figure*}[!htbp]
    \centering
    \begin{tabular}{cc}
\includegraphics[scale=0.55, angle = 0]{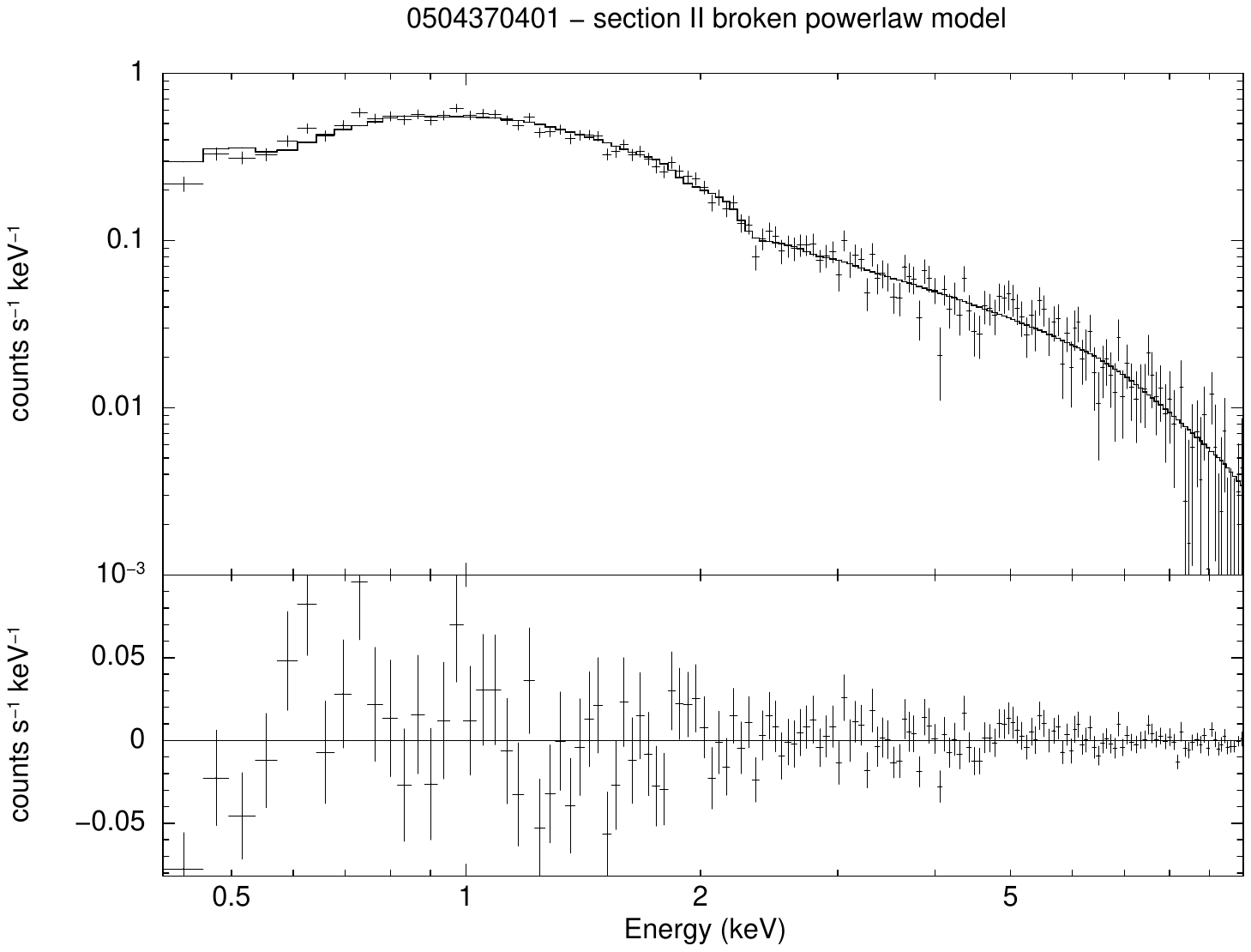} \\ 
\includegraphics[scale=0.55, angle = 0]{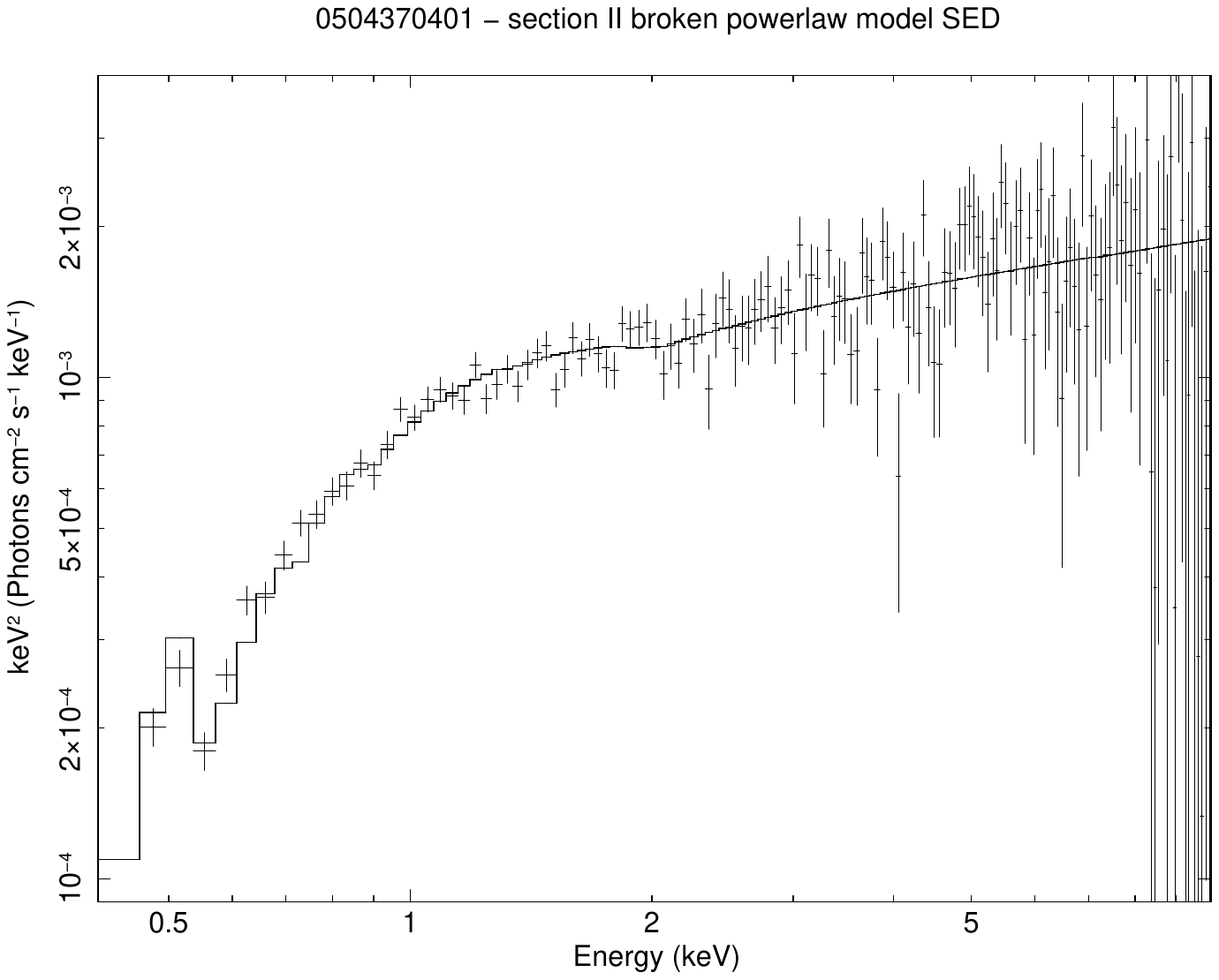} \\
\\
\end{tabular}
\caption{Same as for figure \ref{050437spec_secI} for section II of observation 0504370401.}
\label{0504370401_II_Sed}
\end{figure*} 

\begin{figure*}[!htbp]
    \centering
    \begin{tabular}{cc}
\includegraphics[scale=0.55, angle = 0]{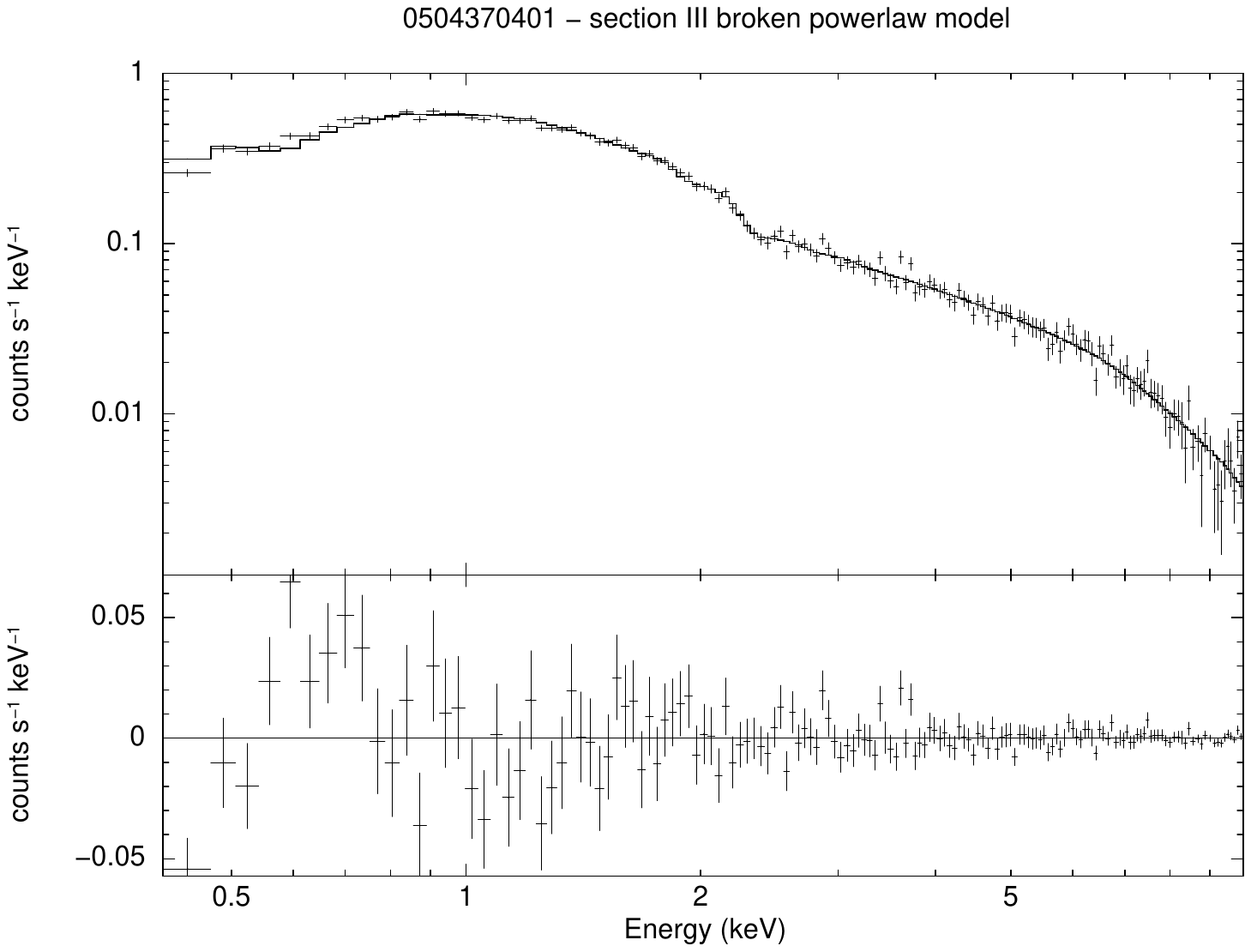} \\
\includegraphics[scale=0.55, angle = 0]{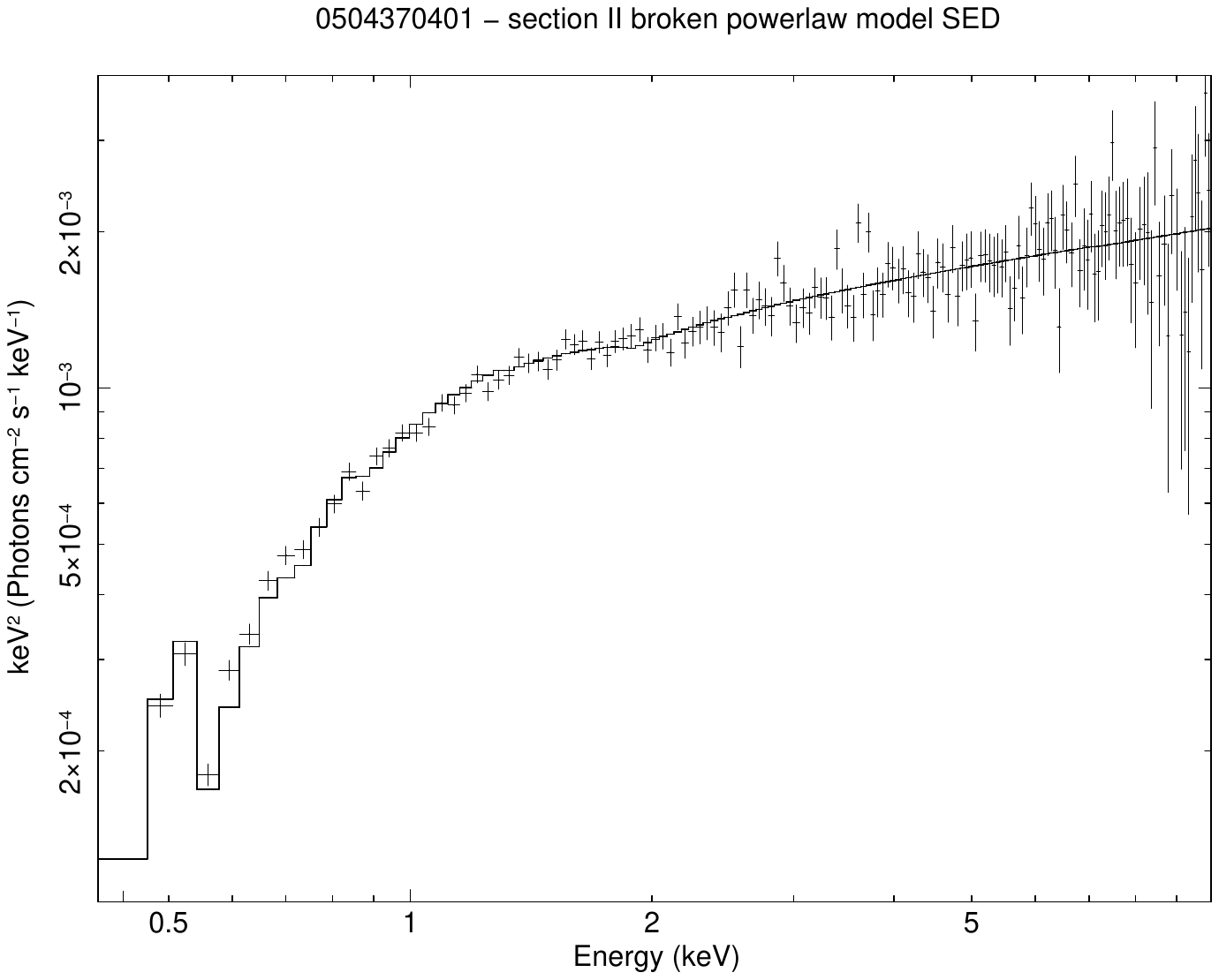} \\
\\
\end{tabular}
\caption{Same as for figure \ref{050437spec_secI} for section III of observation 0504370401.}
\label{0504370401_III_Sed}
\end{figure*} 

\begin{figure*}[!htbp]
    \centering
    \begin{tabular}{cc}
\includegraphics[scale=0.55, angle = 0]{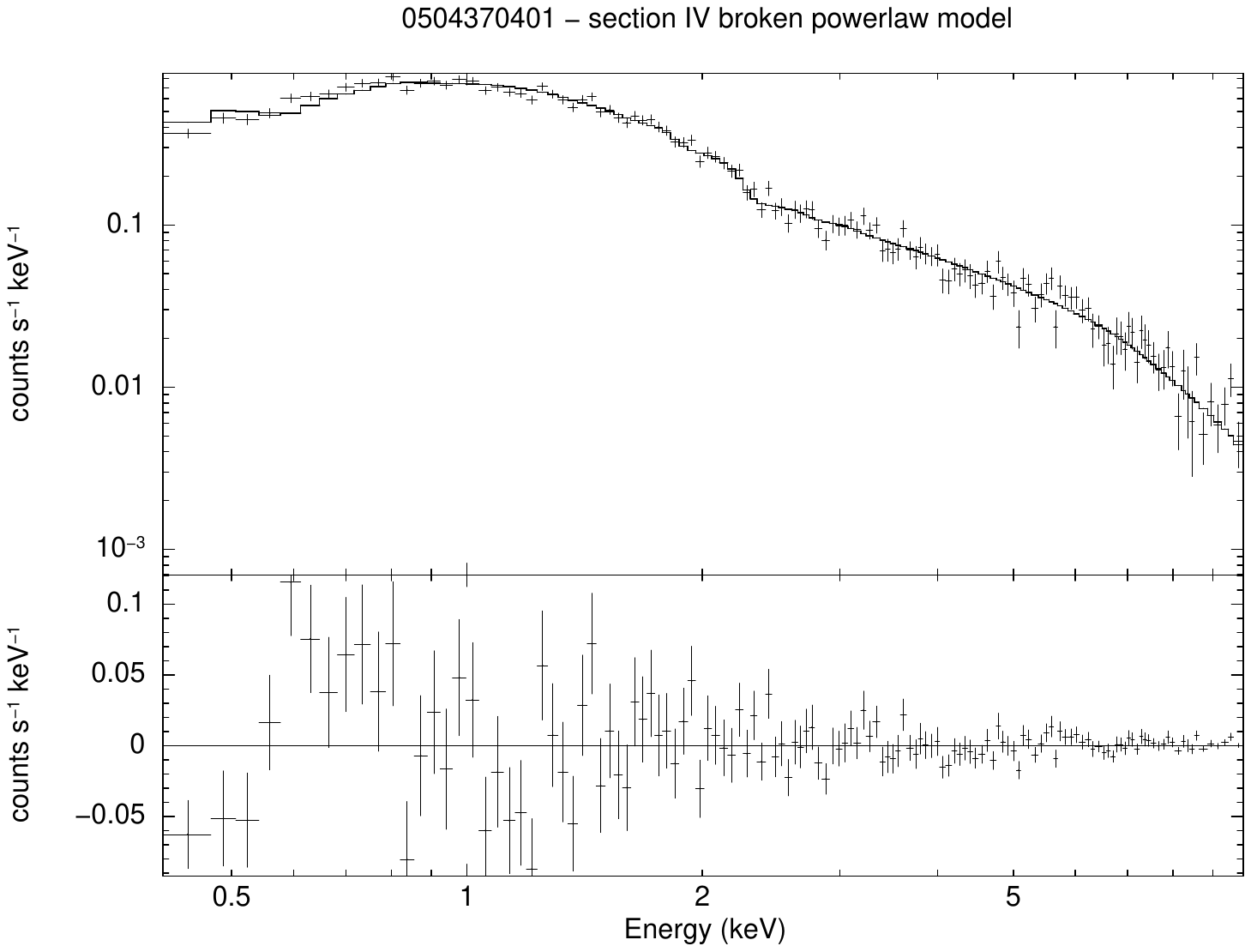} \\ 
\includegraphics[scale=0.55, angle = 0]{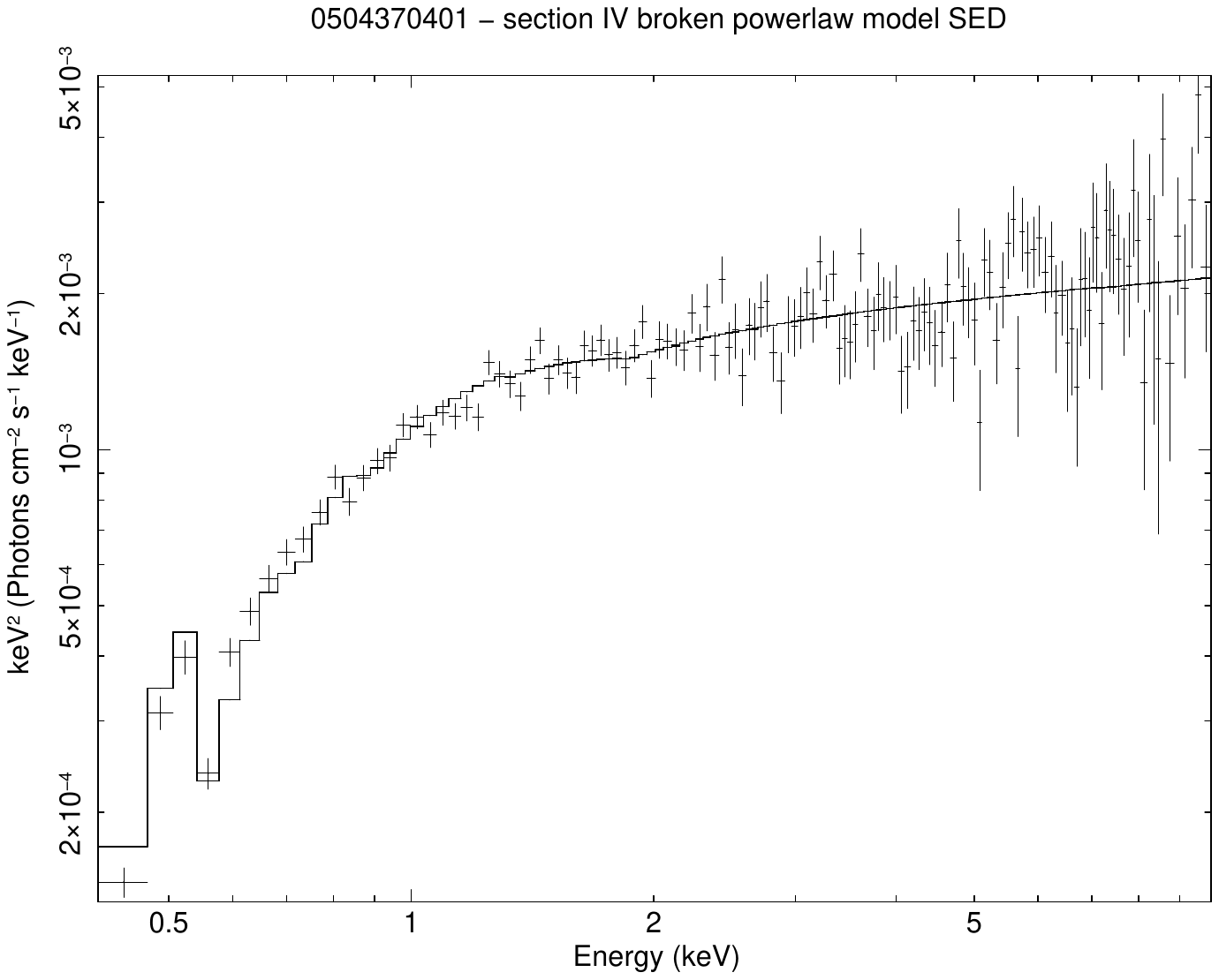} \\ 
 \\
\end{tabular}
\caption{Same as for figure \ref{050437spec_secI} for section IV of observation 0504370401.}
\label{0504370401_IV_Sed}
\end{figure*} 

\begin{figure*}[!htbp]
    \centering
    \begin{tabular}{cc}
\includegraphics[scale=0.55, angle = 0]{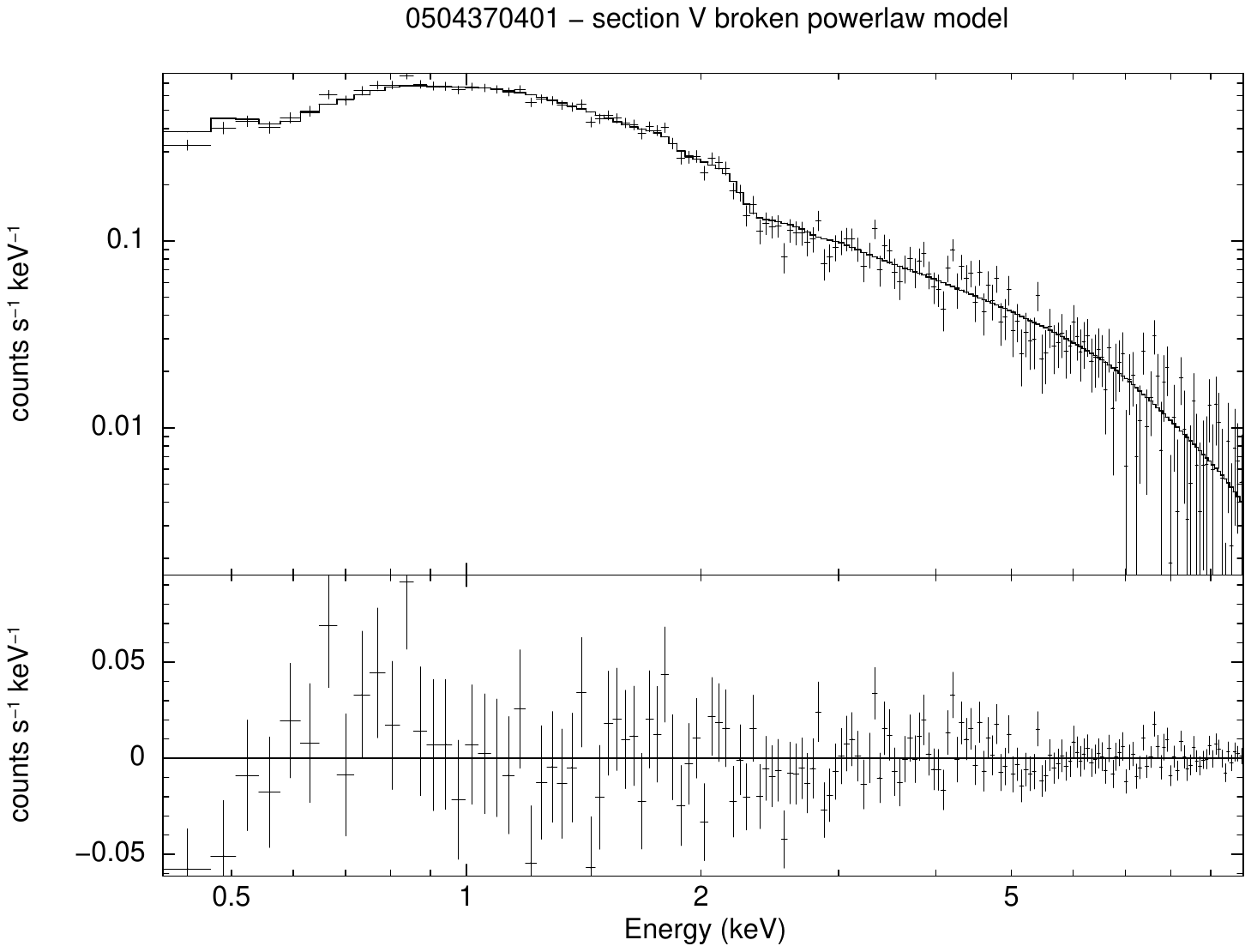} \\
\includegraphics[scale=0.55, angle = 0]{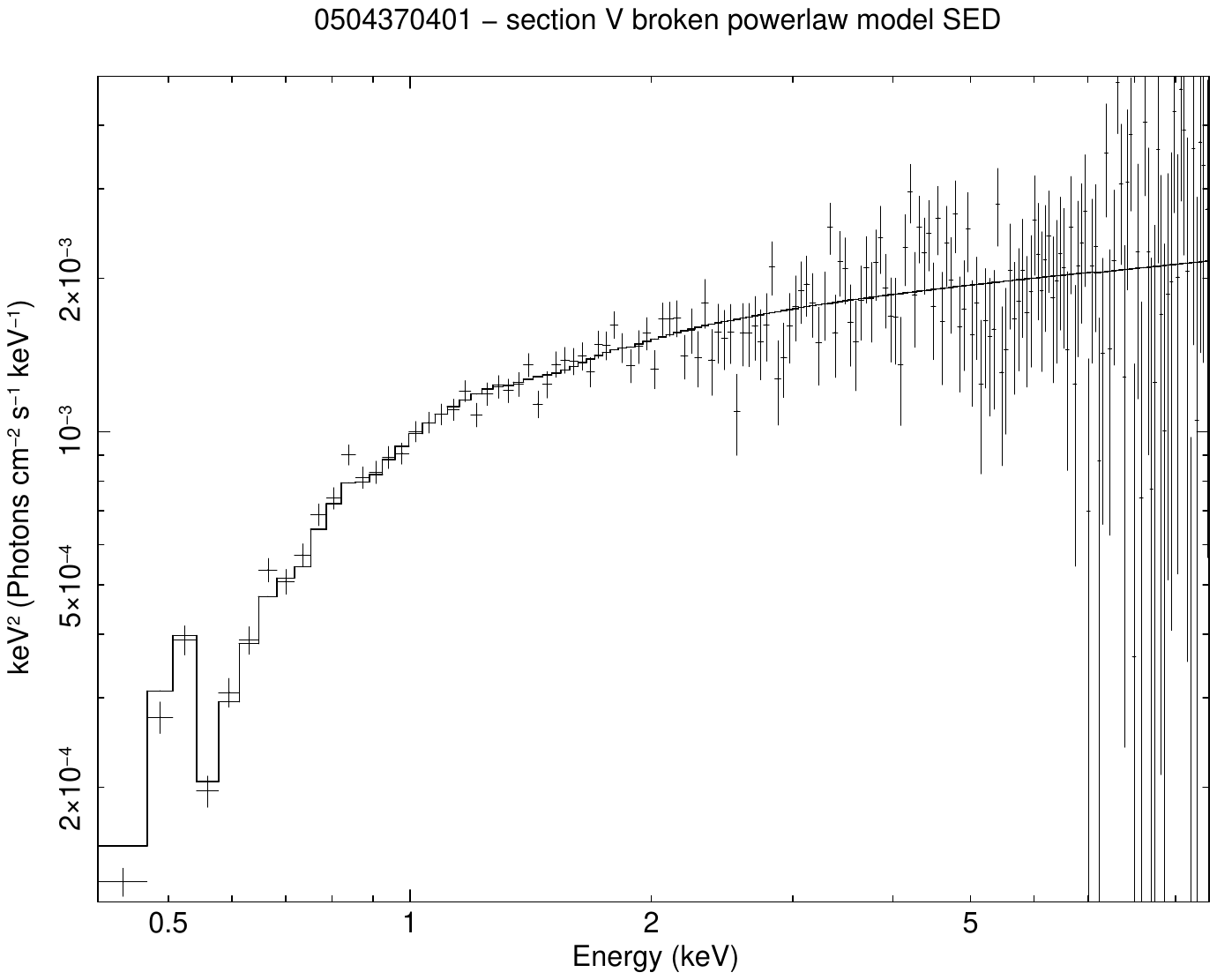} \\
 \\
\end{tabular}
\caption{Same as for figure \ref{050437spec_secI} for section V of observation 0504370401.}
\label{0504370401_V_Sed}
\end{figure*} 

\begin{figure*}[!htbp]
    \centering
    \begin{tabular}{cc}
\includegraphics[scale=0.55, angle = 0]{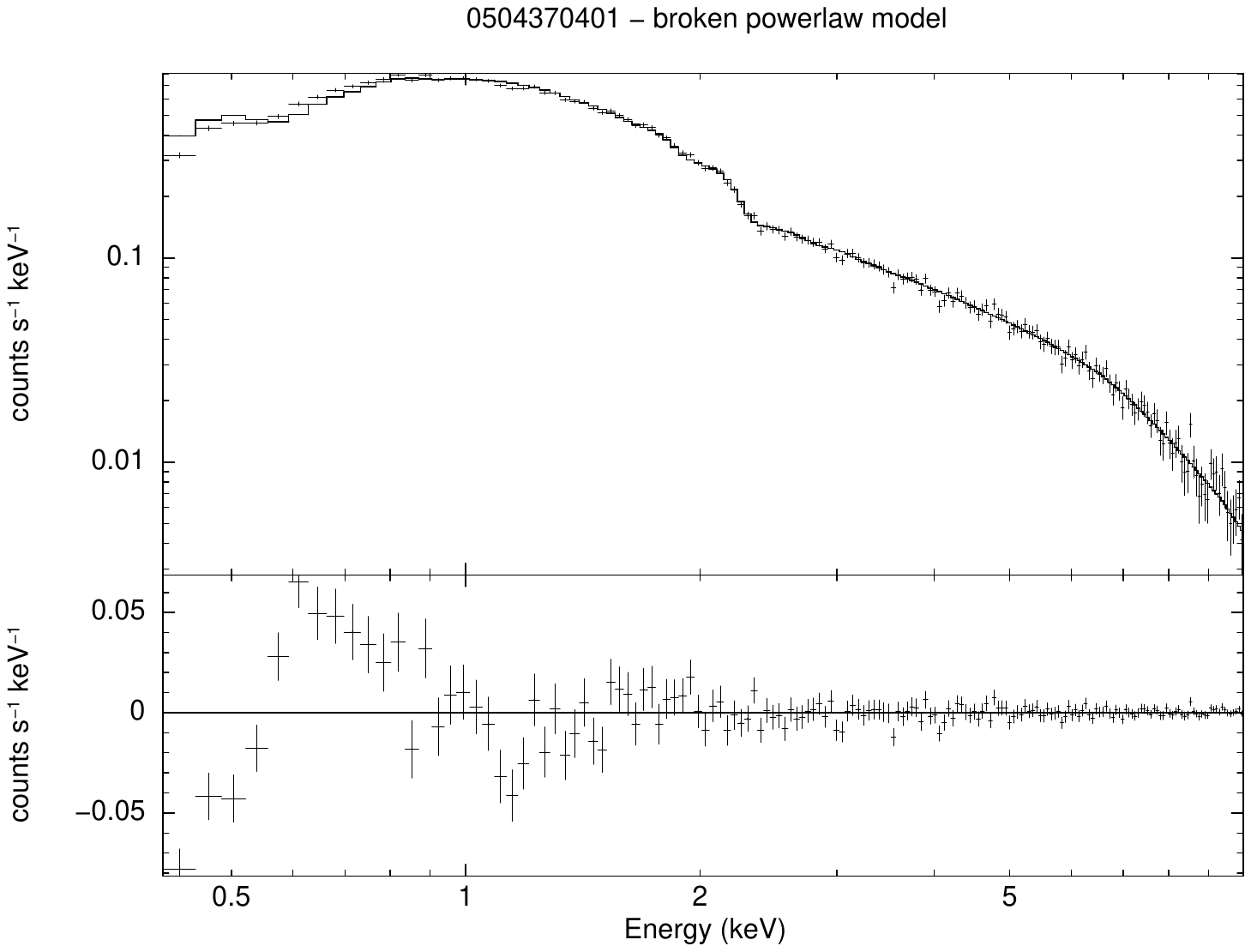} \\
\includegraphics[scale=0.55, angle = 0]{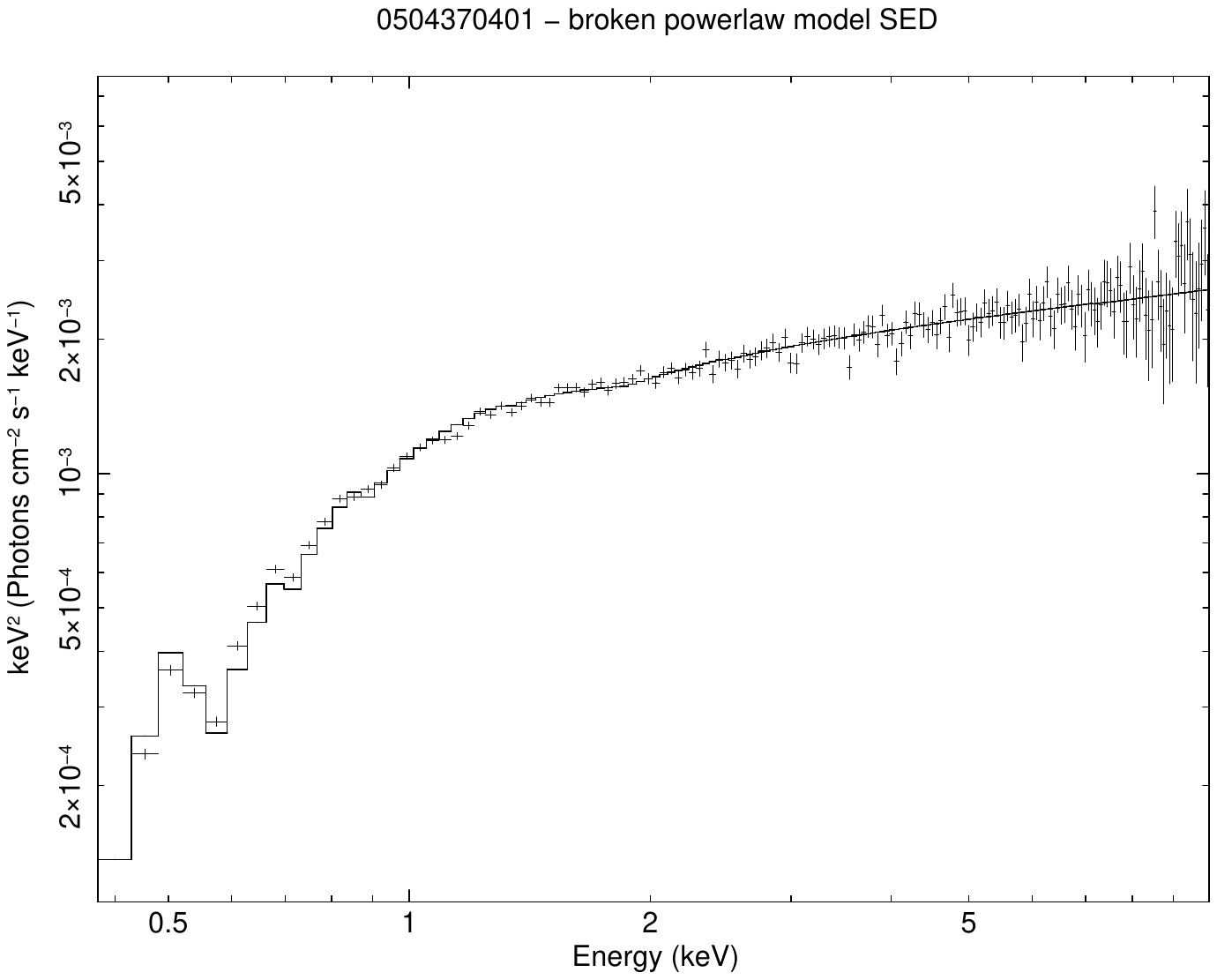} \\
\\
\end{tabular}
\caption{Same as for figure \ref{050437spec_secI} for entire observation 0504370401.}
\label{0504370401_Sed}
\end{figure*} 

\begin{figure*}[!htbp]
    \centering
    \begin{tabular}{c}
\includegraphics[scale=0.55, angle = 0]{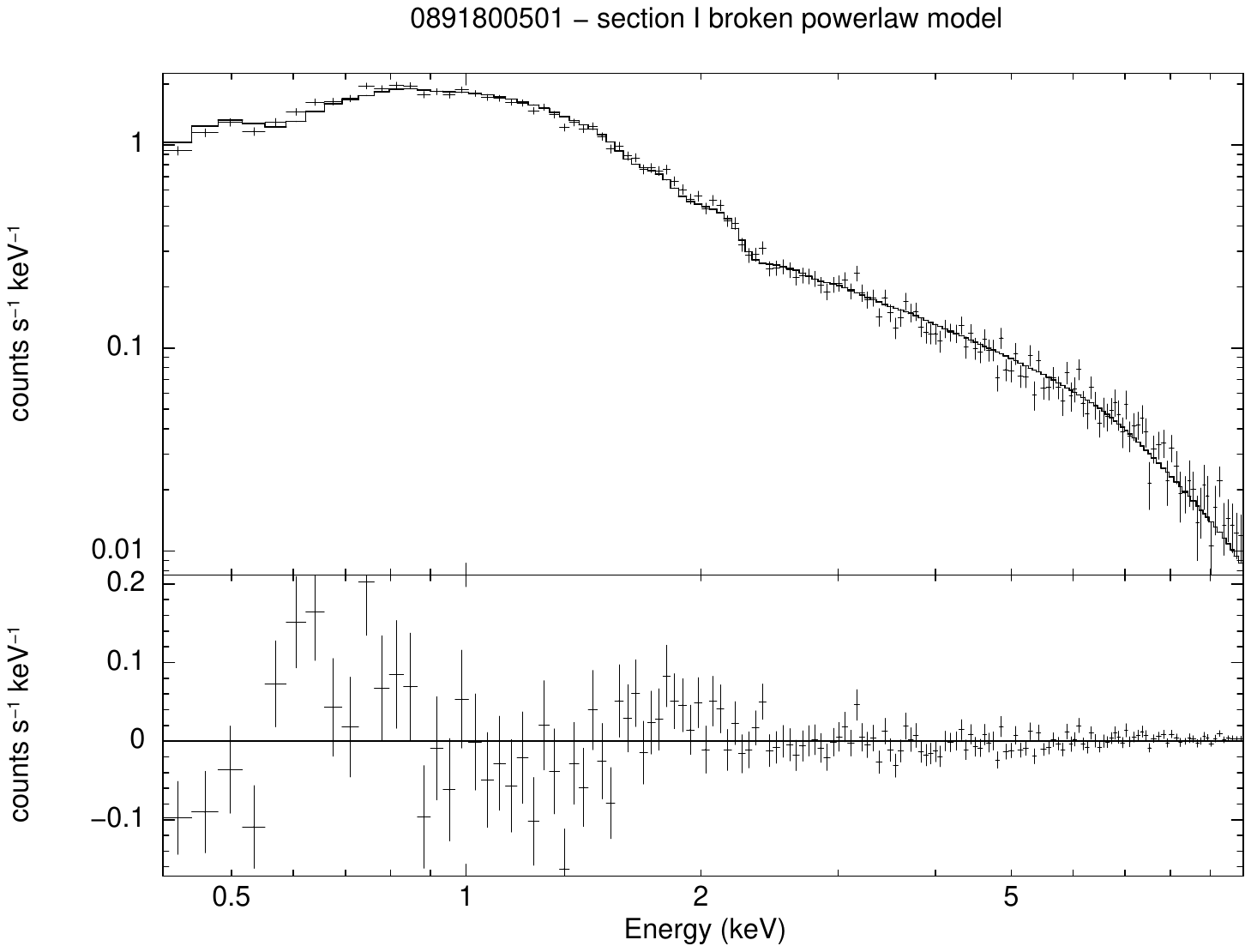} \\
\includegraphics[scale=0.55, angle = 0]{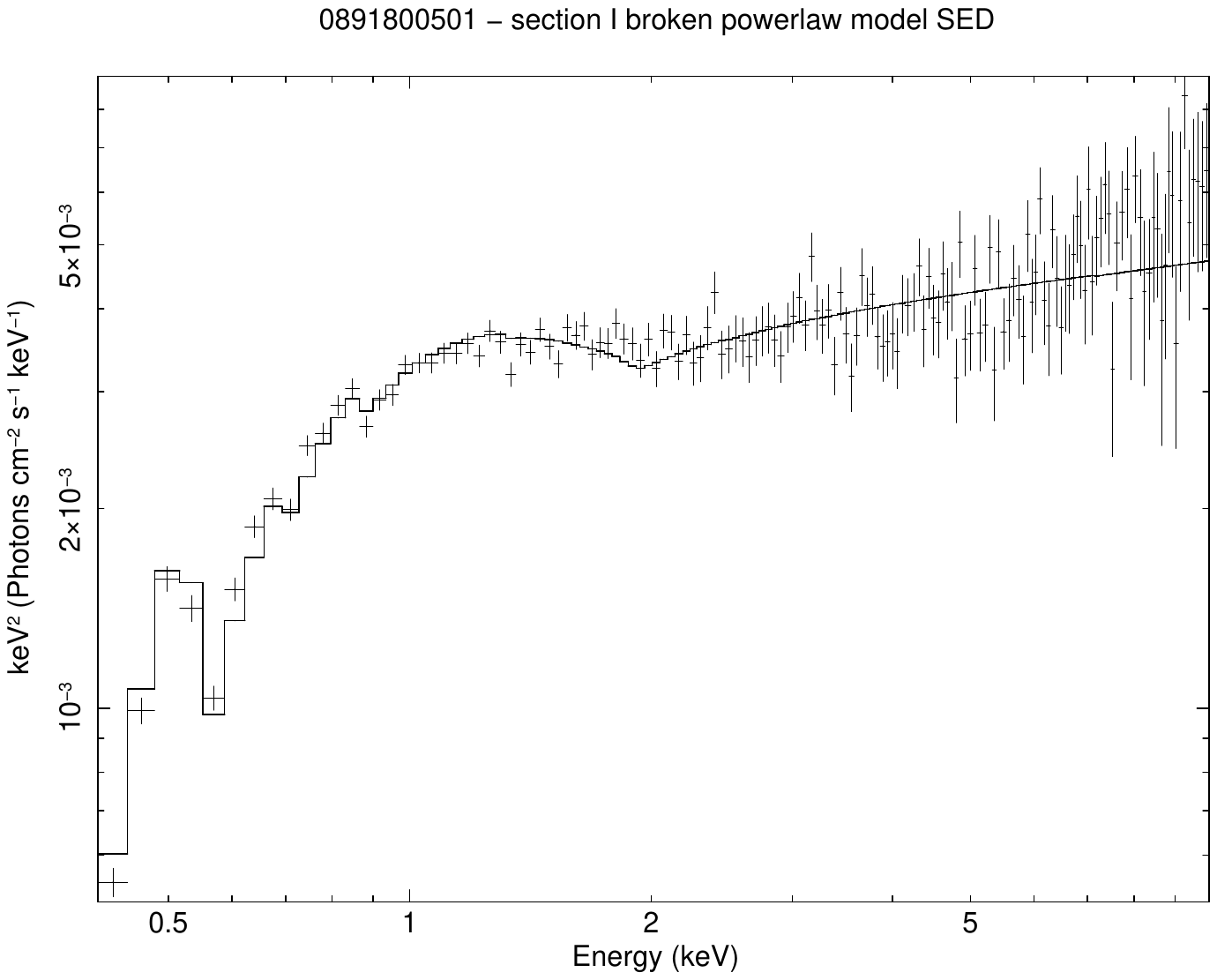} \\  
\\
\end{tabular}
\caption{Same as for figure \ref{050437spec_secI} for section I of observation 0891800501.}
\label{0891800501_I_spectra}
\end{figure*}  

\begin{figure*}[!htbp]
    \centering
    \begin{tabular}{c}
\includegraphics[scale=0.55, angle = 0]{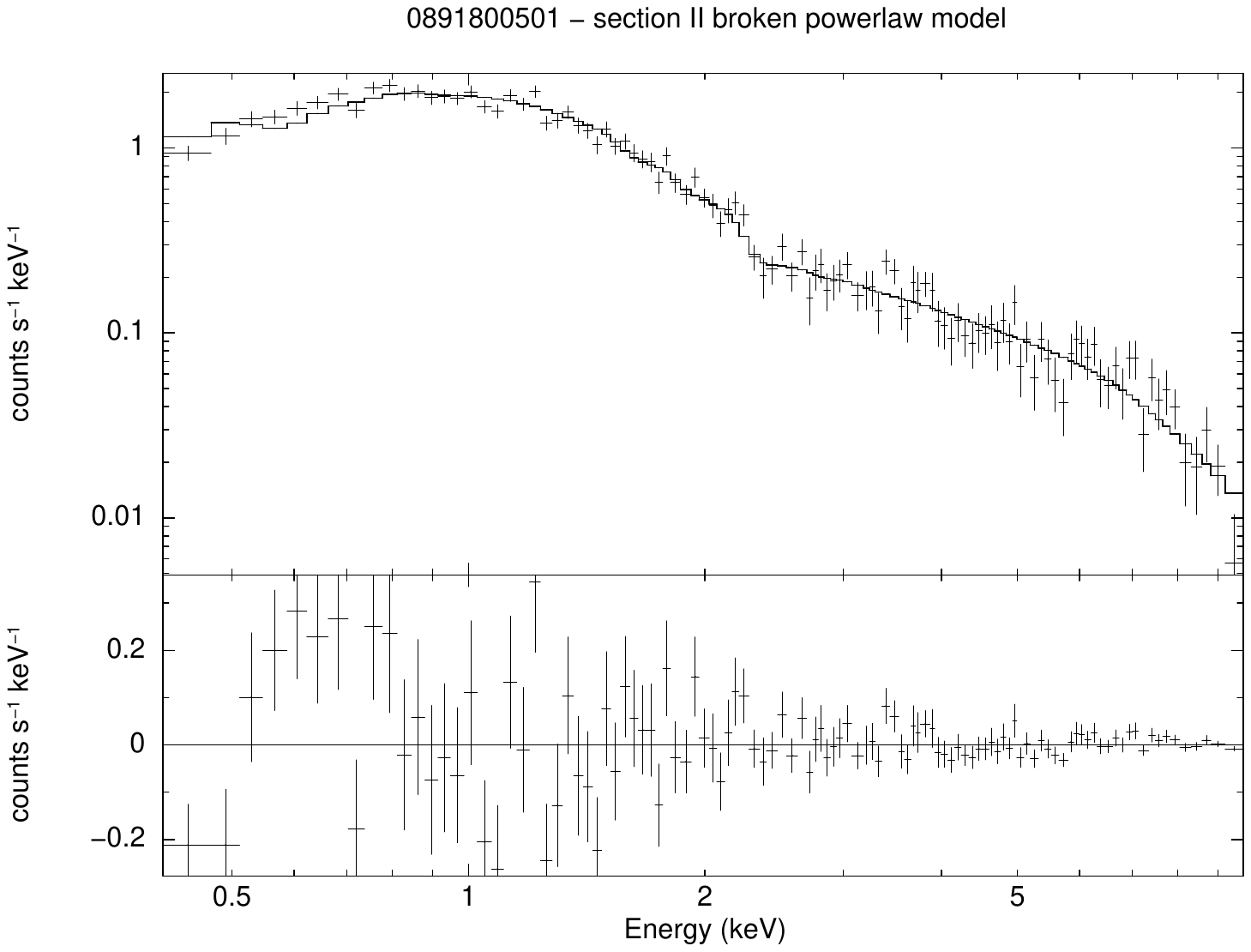} \\
\includegraphics[scale=0.55, angle = 0]{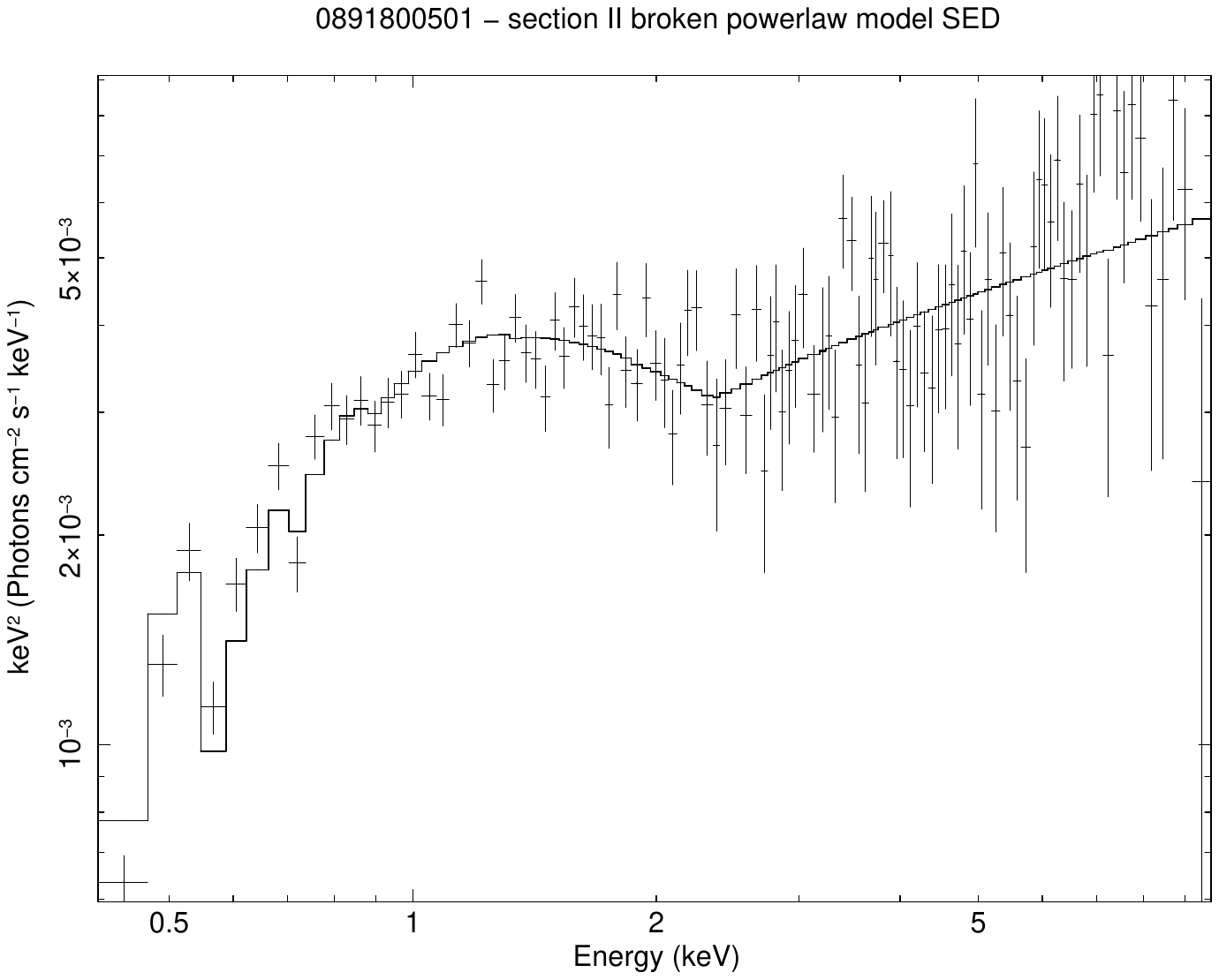} \\  
\\
\end{tabular}
\caption{Same as for figure \ref{050437spec_secI} for section II of observation 0891800501.}
\label{0891800501_II_spectra}
\end{figure*}

\begin{figure*}[!htbp]
    \centering
    \begin{tabular}{c}
\includegraphics[scale=0.55, angle = 0]{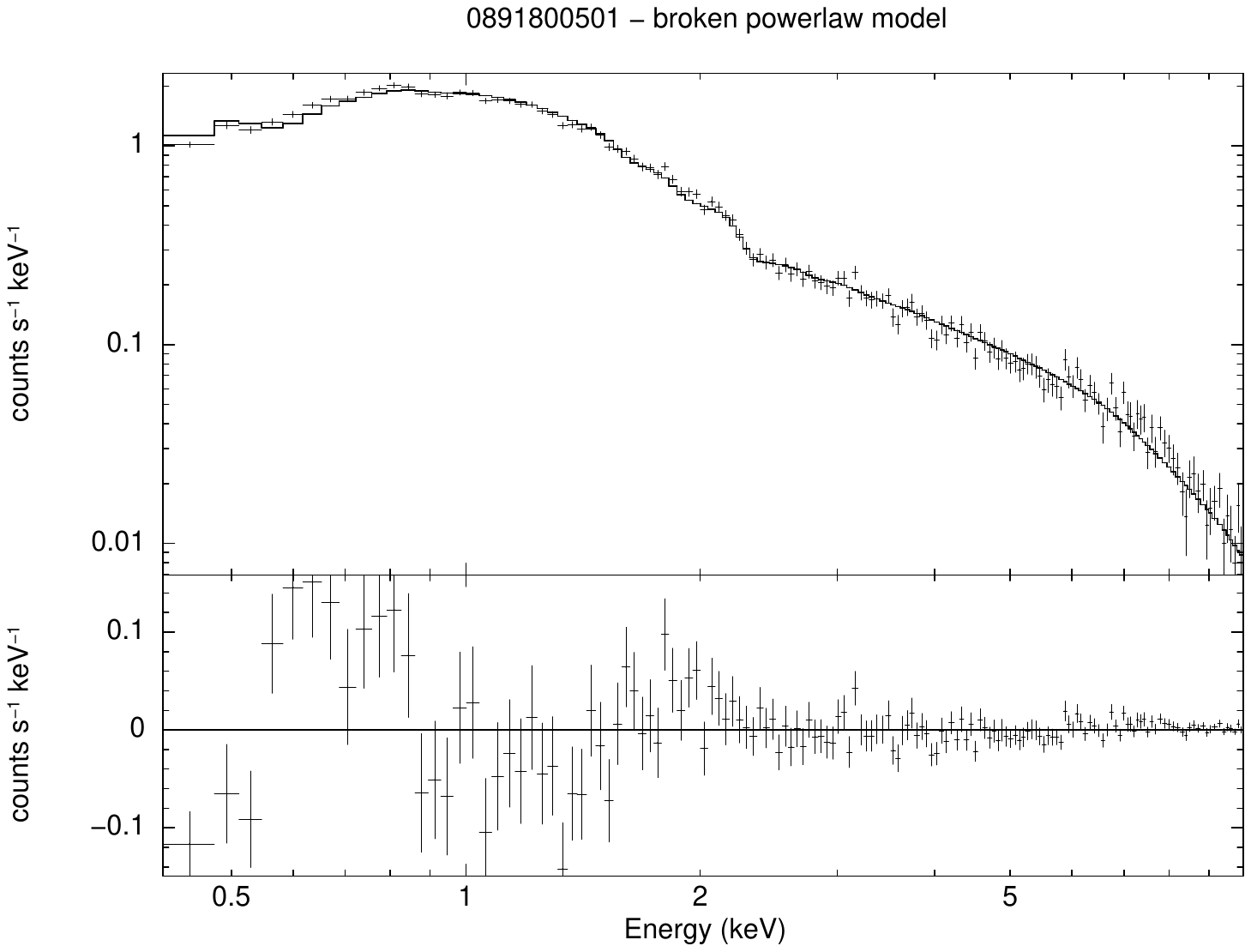} \\
\includegraphics[scale=0.55, angle = 0]{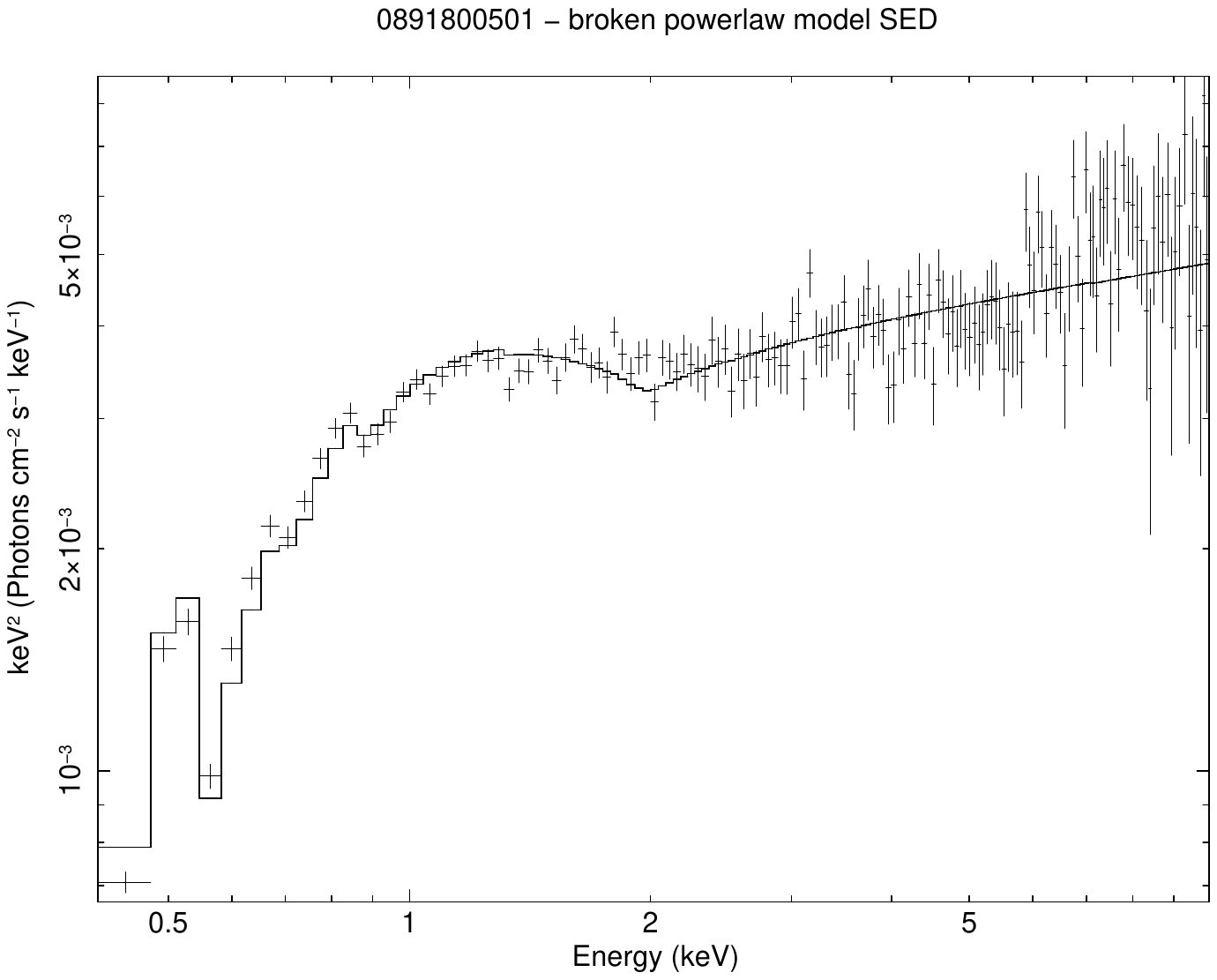} \\ 
\\
\end{tabular}
\caption{Same as for figure \ref{050437spec_secI} for entire observation 0891800501.}
\label{0891800501_III_spectra}
\end{figure*}

\end{document}